\documentclass[a4page,10pt]{article}

\usepackage{fullpage}
\usepackage{amsmath}
\usepackage{amsfonts}
\usepackage{graphicx}
\usepackage{wrapfig}
\usepackage[toc,page]{appendix}

\newtheorem{defn}{Definition}[section]
\newtheorem{lemm}{Lemma}[section]

\author{Puneet Agarwal$^{12}$ \\ puneet.a@tcs.com
\and
Maya Ramanath$^1$ \\ ramanath@cse.iitd.ac.in
\and
Gautam Shroff$^2$ \\ gautam.shroff@tcs.com
\and
Indian Institute of Technology, Delhi, India$^1$ \\
TCS Research, New Delhi, India$^2$}

\date{}

\title{Relationship Queries on Large graphs using Pregel}

\begin{document}
\maketitle

\begin{abstract}
Large-scale graph-structured data arising from social networks, databases, knowledge bases, web graphs, etc. is now available for analysis
and mining. Graph-mining often involves ``relationship queries'', which seek a ranked list of interesting interconnections among a given set of entities,
corresponding to nodes in the graph. While relationship queries have been studied for many years, using various
terminologies, e.g., keyword-search, Steiner-tree in a graph etc., the solutions proposed in the literature so far 
have not focused on scaling relationship queries to large graphs having billions of nodes and edges, such are now publicly available in the form of
`linked-open-data'. In this paper, we present an algorithm for distributed keyword search (DKS) on large graphs, based on the graph-parallel
computing paradigm Pregel. We also present an analytical proof that our algorithm produces an optimally ranked list of answers if run to completion.
Even if terminated early, our algorithm produces approximate answers along with bounds. We describe an optimized implementation
of our DKS algorithm along with time-complexity analysis. Finally, we report and analyze experiments using an
implementation of DKS on Giraph the graph-parallel computing framework based on Pregel, and demonstrate that we can efficiently process 
relationship queries on large-scale subsets of linked-open-data.
\end{abstract}

\section{Introduction and Motivation}
\label{sec:intro}
Many applications produce or deal with large graphs. Such graphs could be entity-relationship graphs extracted from textual sources, relational databases modelled as graphs using foreign-key relationships among tuples, biological networks, social networks, or call data records\footnote{The graph capturing data about people calling each other.}. Large graphs encountered in practice are both node-labeled (containing entity description) as well as edge-labeled (indicating semantics of relationship between nodes). Based on such labels, numeric weights can be assigned to the edges of the graph, if they are not already available, which indicate strength of relationship between the corresponding nodes. Since graphs generated by web-scale social applications are often massive, efficiently querying and analyzing them is a non-trivial exercise.
 
While graphs can be queried in a variety of ways, we are interested in a specific class of queries called
\emph{relationship queries} \cite{p:dks:cikm}. Here, a set of entity names are given as query keywords, and the objective is to find a node (\textit{root-node}) in the graph such that the nodes that represent the given entities are connected to the root-node with a shortest path. 

\begin{wrapfigure}[14]{R}{0.45\columnwidth}
\vspace{-15pt}
\centering
    \includegraphics[width=0.45\columnwidth]{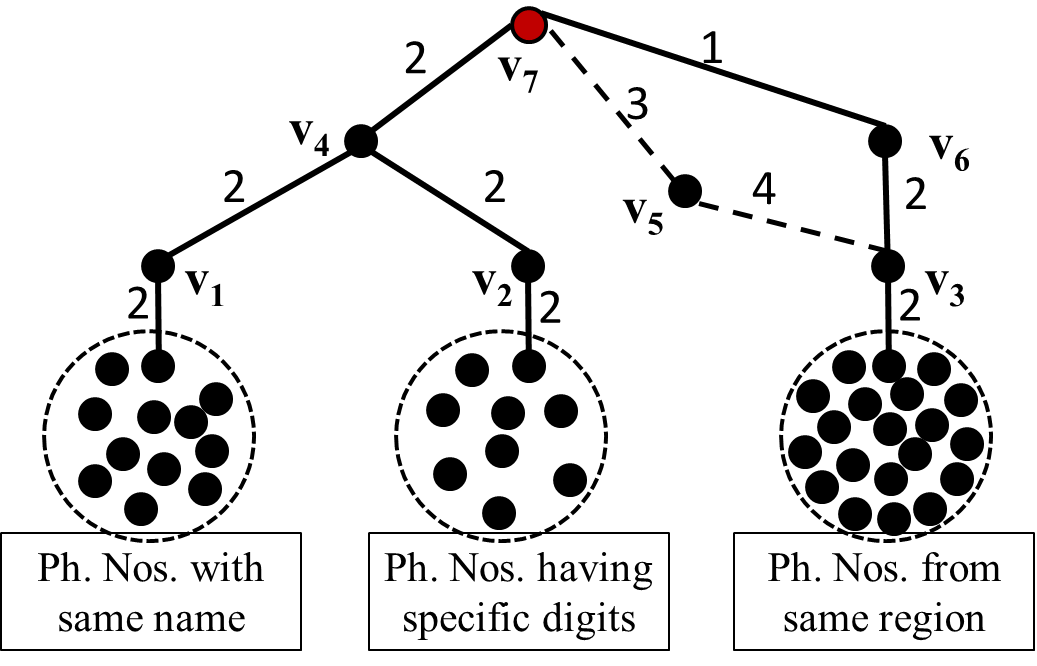}
   \caption{Example Relationship Query}
  \label{fig:intro}
\end{wrapfigure}

Relationship queries are particularly useful while mining for complex graph patterns, such as the detection of collusive frauds, where we want to discover relationships between entities, which should not exist normally. Consider an agency that has leads from multiple terrorist activities such as phone-numbers of people involved in acts of terrorism. They often want to discover whether there is any relationship between those leads, and identify a node (root-node) which connects them all. 
For example, if there are three leads, i)~cellphones operating from a specific region, ii)~specific digits in phone numbers, and iii)~cellphones of people with specific names, as shown in Figure \ref{fig:intro}. This gives us three groups of nodes of call data record graph, and the \textit{relationship queries} can be used to find interesting interconnections among three nodes (one node from every group). In our example, $v7$ is the connecting node between these leads, and the answer-tree is shown in Figure \ref{fig:intro}. We aim at finding top-K answers-trees, such that the edges in the answer-tree have small weight.

Another use-case that can be addressed using relationship queries is detecting \textit{Insider Trading}, i.e., trading of stocks of a company by taking a cue from insider (non-public) information about the company. Insider trading is illegal in most of the countries. Relationship queries on a graph, constructed from a database of key office bearers of companies and their relationships with other key people, can be used to discover instances of insider trading. In a similar manner, investigating agencies might want to discover relationships between politicians and industrialists, whom they favor. A form of such relationship queries is also used for identifying K-effectors in a social media network \cite{p:MannilaKeff}.

The problem of relationship queries is very similar to that of keyword search on graph data \cite{p:banks1}. Prior work on keyword search on graphs has focused on standalone algorithms \cite{p:banks1} where the graph is small or has advocated the use of pre-processing of graphs \cite{p:banks3} and the use of indexes \cite{p:blinks} to overcome the memory bottleneck. Most of these approaches don't even make an attempt to find the optimal answer. Further, these solutions cannot scale and a distributed algorithm is indicated. Similar to the related work, we also present the answer of a relationship query in the form of a tree, called answer-tree.

\textbf{\textit{Key Contributions}:} In this paper, we present \textit{\textbf{i)}}~an algorithm called \textit{distributed keyword search} (DKS) for the problem of relationship queries, which is based on distributed parallel processing techniques. Finding an answer of such queries in the form of a tree such that it contains all query entities, and has the smallest answer-tree weight is considered hard. In DKS, we search for the root-node of the answer-tree following in-parallel breadth first search (BFS) approach, and find the optimal answer most of the time. For some queries, completion of BFS may take long-time and it may keep searching for optimal answer-tree forever. We therefore do not propose the traversal of entire graph and stop running subsequent iterations of BFS after DKS' exit criterion is satisfied. 
\textbf{\textit{ii)~}}We also present an analytical proof for optimality of answers discovered by DKS, i.e., when exit criterion is satisfied and further iterations of BFS are stopped, the optimal answer is not missed. Further, sometimes BFS may become extremely slow before the exit criterion is satisfied. 
\textit{\textbf{iii)~}}In such situations also, we stop further iterations of BFS, and estimate the degree of optimality of the answers discovered so far. 
\textbf{\textit{iv)~}}We also present an optimization of basic approach for finding a locally optimal answer-tree at every node visited during BFS traversal of the graph, as part of DKS. 
\textit{\textbf{v)~}}We analyze the time-complexity of DKS and demonstrate that the time taken by DKS is linear in size of the graph, however exponential in the number of keywords, which is an acceptable norm for relationship queries. 
\textbf{\textit{vi)~}}Arguments made during the description of algorithm and analytical proof, are finally corroborated by presenting empirical benchmarks on largest ever dataset attempted by baseline approaches for this problem.

\textbf{\textit{Organization of the Paper}}: We begin with a formal description of the problem in the next section, and present a brief description of the related work in Section \ref{sec:rw}. We then describe the DKS algorithm in Section \ref{sec:overview}, and in Section \ref{sec:opti} we first present an optimization of the basic DKS algorithm and then analyze the time-complexity. The analytical proof for optimality of the answers discovered by DKS has been included in Section \ref{sec:proof}. Next, we share performance benchmarks of DKS on largest ever graph data reported in related research publications, in Section \ref{sec:exp}. Finally, we conclude in Section \ref{sec:conslusion}.

%================================================
\section{Problem Desc. \& Definitions}
\label{sec:probDesc}
Our definition of the problem about \textit{relationship queries} is motivated by the prior work \cite{p:banks1,p:banks3,p:blinks,p:banks_ease}. We assume that the input data is represented as a graph $G = (V,E)$. Here, $V$ is the set of nodes and $E$ is the set of directed edges between pairs of nodes. Every node $v \in V$ has associated text such as name of an entity, and every edge $e \in E$ has associated label as well as positive numeric weights, i.e., $0 ~<~w(e) \in \mathbb{R}, \forall e\in E$. The label on an edge provides semantic information about the relationship between its end nodes, and reciprocal of the numeric edge weight represents the strength of this relationship. Therefore for better intuition the edges weights are referred to as edge-lengths.

The objective is to execute a relationship query, comprising of a set of keywords $Q = \{q_1, q_2, ..., q_m\}$, on the graph $G$. Here, keywords can be names of entities, and nodes that contain any keyword of the query are called keyword-nodes $t_i^j$. Here, $T_i = \{t_i^1, t_i^2, ...,\}$ is the set of keyword-nodes containing keyword $q_i$. The answer $A$ of such queries is presented as a tree, which is a subset of the graph $G$ such that nodes of the answer-tree collectively contain all keywords of the query. The answer is represented as a tree because the root-node of such a tree is the common connection between all keywords (entities) of the query and finding such a node is the key objective of relationship queries as explained earlier.

% \newdef{definition}{Definition}
\begin{defn}
\label{def:ans}
In the context of relationship queries on a graph $G$, a \textit{minimal answer-tree} $A \subseteq G$ is defined as a tree, such that the nodes of the tree collectively contain all keywords of the query and by removing any node/edge from the answer-tree the remaining data-structure does not remain connected or does not contain all keywords of the query.
\end{defn}

Here, the \textit{weight of an answer-tree} $w(A)$ is calculated as sum of the lengths of all the edges of the tree, i.e., $w(A) =\sum_{e\in A} w(e)$. Since the reciprocal of edge-length indicates strength of relationship between the pair of nodes, the answer-weight should be as small as possible. The problem of relationship queries is defined below.

\begin{defn}
\label{def:rq}
Given a set of keywords $Q = \{q_1, q_2, ..., q_m\}$, in the context of a graph $G$, find the $K$ best minimal answer-trees, in increasing order of their weights.
\end{defn}

It can be observed that the above problem is equivalent to the Group Steiner Tree (GST) problem, as also shown in \cite{p:DPBF,p:sagiv1}. The GST problem is defined as: \textit{given a set of $m$ groups of nodes of graph $G$, find a minimal spanning tree, such that the tree contains at-least one node from every group of the nodes} \cite{p:steinerSurvey2013}. In case of relationship queries, keyword-nodes $T_i$ of every keyword $q_i$ are equivalent to a group of nodes of the GST problem, and the minimal answer-tree is equivalent to minimum spanning tree.

%================================================
\section{Related Work}
\label{sec:rw}
Most of the prior work \cite{p:kwdSrchSurvey} on this problem propose a standalone solution for generating heuristic solutions. Many of these algorithms \cite{p:banks1,p:banks3,p:banks2} don't even measure the degree of approximation of the answers produced by their algorithm, since the problem is hard. While, we either generate an optimal answer or predicts the degree of approximation using DKS. This problem has many different interesting aspects such as ranking of the answers discovered \cite{p:dbXplore,p:Objectrank1,p:banks1,p:xrank}, and different possible structures of the answers \cite{p:rCliquesKS} itself. Many different methods have been proposed as a solution to this problem, such as graph traversal\cite{p:banks1}, SQL queries\cite{p:Discover,p:dbXplore}, and clustering and index guided methods\cite{p:blinks,p:banks_ease}. DKS follows graph traversal based method.

The Steiner Tree problem on graphs was surveyed by Bezensek et al. in \cite{p:steinerSurvey2013}. According to this and other such surveys most of the researchers have been trying to find a heuristic solution to this problem, such as  Shortest Path Heuristic, Average Distance Heuristic, and Distance Network Heuristic etc. Most of these have an approximation ratio $\theta=2$; here, $\theta$ is ratio of approximate answer weight detected by an algorithm and the optimal answer weight. By and large the best solution was presented by Robins et al. \cite{p:SteinerSoln2000} with 1.55-approximation guarantee. To the best of our understanding there have been no effort on trying to restrict the search space of this problem, which is one of the primary contribution of our work. Kimelfeld et al. in \cite{p:sagiv1} highlighted that according to \cite{p:steiner,p:STPSmallK}, the Steiner Tree problem is solvable with bounded number of keywords, and they also present a heuristic approach. We also corroborate the same finding using time-complexity analysis of our algorithm in Section~\ref{sec:tc}.

The Steiner Tree problem or Group Steiner Tree problem has been attempted in multiple domains such as for routing of network packets in computer networks, multiple applications in social networks\cite{p:SocTeamSteiner,p:MannilaKeff}, identification of functional modules in protein networks\cite{p:PPI}. Most of these algorithms are either a heuristic approach or apply a domain specific constraint to solve this problem.

Many a heuristic solutions using distributed and parallel computing for this computationally intractable problem have been presented and were surveyed in \cite{p:steinerSurvey2013}. The solutions that use parallel processing paradigm are based on a shared memory across all the processors, such as the hybrid genetic algorithm based approach proposed by Presti et al. \cite{p:parallelSTP}. Bauer et al. \cite{p:distSTP1,p:steinerSurvey2013} presented distributed algorithms based on K-SPH (Kruskal’s shortest path heuristics)[6]. Here, keyword-nodes with lowest index are called terminal leaders. The leaders of a sub-tree are made responsible for the co-ordination of subtree. The closest sub-trees are merged using discovery and connection steps. Later Singh et al. \cite{p:distSTP2}, improved the in-efficiencies of the discovery step and presented a solution which performed better than its original variant. An important limitation of such approaches is that they cannot work on large graphs and is not based on modern parallel processing paradigm. Recently, \cite{p:kwdSrch:mr} presented a solution to the keyword search problem using map-reduce\cite{p:mr} paradigm, we argue that Pregel is better choice of distributed processing paradigm for this problem, since it does need to load the entire graph in every iteration/superstep. We \cite{p:dks:cikm} present a solution to this problem using parallel processing paradigm Pregel\cite{p:pregel}, which either discovers an optimal solution or a heuristic answer along with its approximation ratio.

%================================================
\section{Background \& DKS Overview}
\label{sec:overview}
Our distributed keyword search algorithm makes use of Pregel\cite{p:pregel} model for distributed processing. Therefore we first provide a brief overview of Pregel.

\textbf{\textit{Pregel Overview}}: Pregel jobs run on a compute-cluster, and every computer can be configured to have more than \textit{worker agents}, which run in parallel and perform most of the work. When it starts to process any job on an input graph, it first distributes every node of the graph to a specific \textit{worker}, chosen using a hash-function. In Pregel framework, the input graph is processed iteratively and these iterations are called \textit{supersteps}. In every superstep, a user-defined \textit{compute()} function gets called for every node of the graph on its worker, independently. Here, one common \textit{compute()} function is defined for all nodes of the graph, and for all supersteps of a Pregel job. Through this \textit{compute()} function, nodes send and receive messages to/from each other. When role of a node in a Pregel job is deemed to have been completed, we call a library function \textit{voteToHalt()} from \textit{compute()} function, which indicates to the Pregel framework that the current vertex will remain dormant from subsequent superstep onwards. Such dormant nodes are referred to as \textit{inactive nodes} and remaining nodes are referred to as \textit{active nodes}. If an \textit{active node} sends a message to \textit{inactive node}, it becomes active again. The processing comes to a halt when all nodes become inactive. 
Further, Pregel framework has a provision for another agent called \textit{Aggregator}, which is a user-defined function that can receives 
messages from all nodes of a superstep, and aggregates the messages sent to it. 
The aggregated value of the messages sent in a superstep $s$, is made available to all nodes in the next superstep $s+1$.

To describe the DKS algorithm, we take help of a few terms that are defined here. A subset of query keywords $k_i \subseteq Q$ is called \textit{keyword-set}. Set of all keyword-sets is the power-set of $Q$, i.e., $\mathcal{P}(Q) = \{k_1, k_2, ..., k_{(2^m-1)}\}$; here, $k_i \ne \phi$. If we drop one or more keyword-nodes and related edges from a minimal answer-tree, the tree that is left is called \textit{partial answer} $A'$. We also define \textit{path-length} $\phi(k_i,v)$ of a keyword-set $k_i$, at a node $v$, as weight of a partial answer rooted at $v$ and containing keyword-set $k_i$. 

\begin{wrapfigure}[26]{R}{0.6\columnwidth}
% \vspace{-15pt}
    \includegraphics[width=0.6\columnwidth]{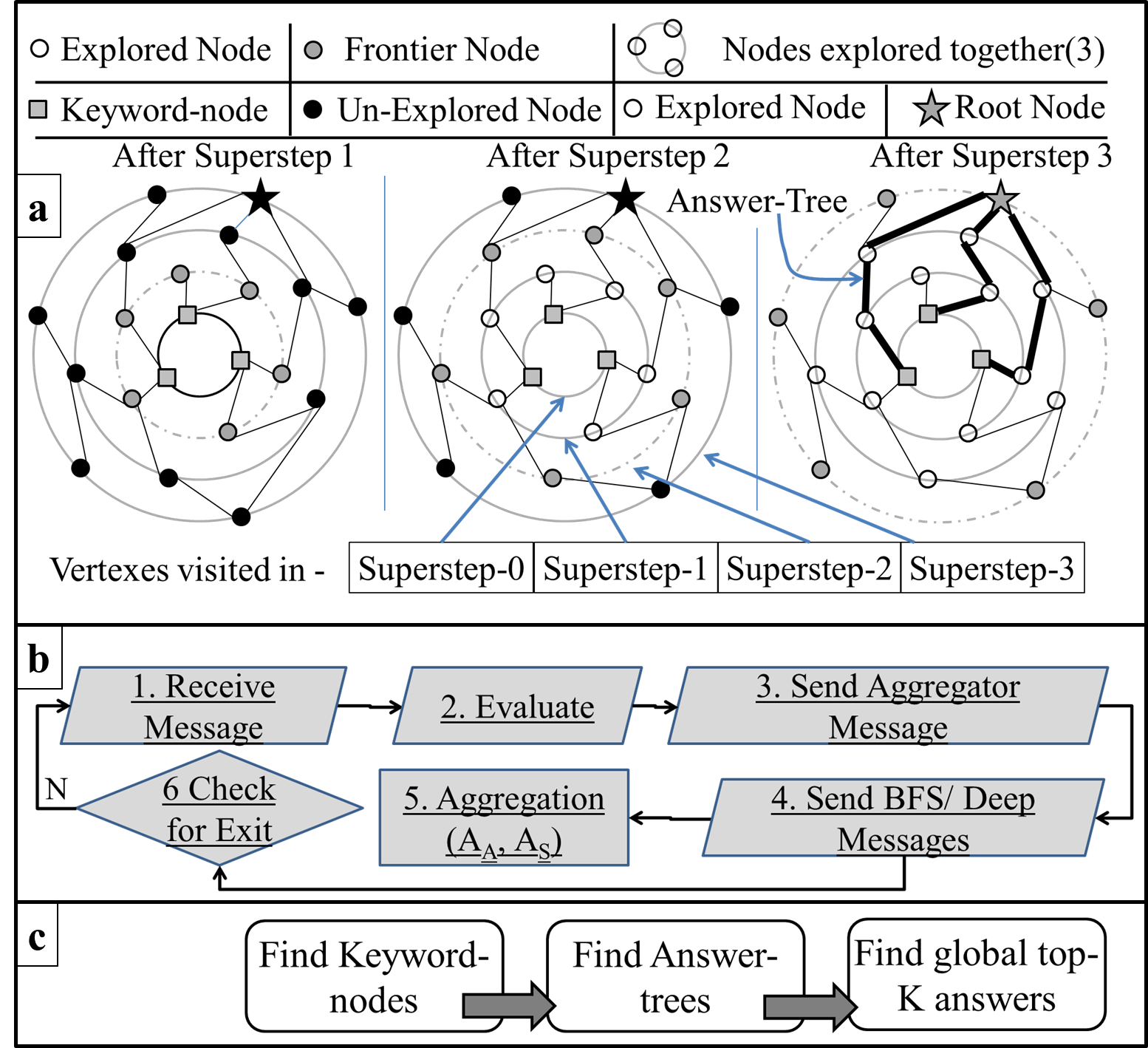}
   \caption{a)~BFS traversal example b)~DKS Flowchart in Pregel c)~High level steps of DKS.}
  \label{fig:bfs}
\end{wrapfigure}

\subsection{DKS Algorithm}
We first pre-process a graph and prepare it for running DKS. Here, we calculate the edge-lengths if not already present and also create an inverted-index \cite{p:invertedIndexSurvey} of the text associated with the nodes of the graph $G$. For all directed edges of the graph, we also include the reverse edges with the same edge-weight so that suitable answer-trees could be discovered irrespective of the direction of relationship between nodes. High-level flow of DKS algorithm is shown in Figure \ref{fig:bfs}(c), here, we first search for the query keywords in the inverted index, and identify the keyword-nodes that become the starting points of parallel BFS traversal (Find Answer-trees). During BFS traversal of the graph, at every node we evaluate whether it has a path to all keyword-nodes of the query. BFS traversal on a contrived example is explained in the next paragraph. All such answer-trees, found at various nodes of the graph, are aggregated to find the global top-K answer-trees.

We explain the BFS traversal of DKS, through a contrived example shown in Figure \ref{fig:bfs}(a). Here, in the first ($0^{th}$) superstep, we send messages from keyword-nodes to their neighboring nodes. The message contains paths to the keyword-node from the neighboring node, and corresponding path-length. All the other nodes of the graph remain dormant. The neighboring nodes of keyword nodes receive the message(s) in the next superstep. Such nodes send a message to their unexplored neighboring nodes. The message contains paths to keyword-nodes known at the sending node. This process continues through subsequent supersteps. The state of a sample graph after superstep-1, 2, and 3 is shown in Figure \ref{fig:bfs}(a). When a node receives a message for the first time, it is declared as \textit{Frontier node}. Finally, in a superstep ($3^{rd}$ in our example), the control reaches a node (star-marked) that has path to at-least one keyword-node of every keywords of the query, i.e., it is the root-node of an answer-tree. A more detailed description of the DKS algorithm, is given below with the help of a flowchart shown 	in Figure \ref{fig:bfs}(b).

\begin{wrapfigure}[23]{R}{0.5\columnwidth}
   \centering
    \includegraphics[width=0.5\columnwidth]{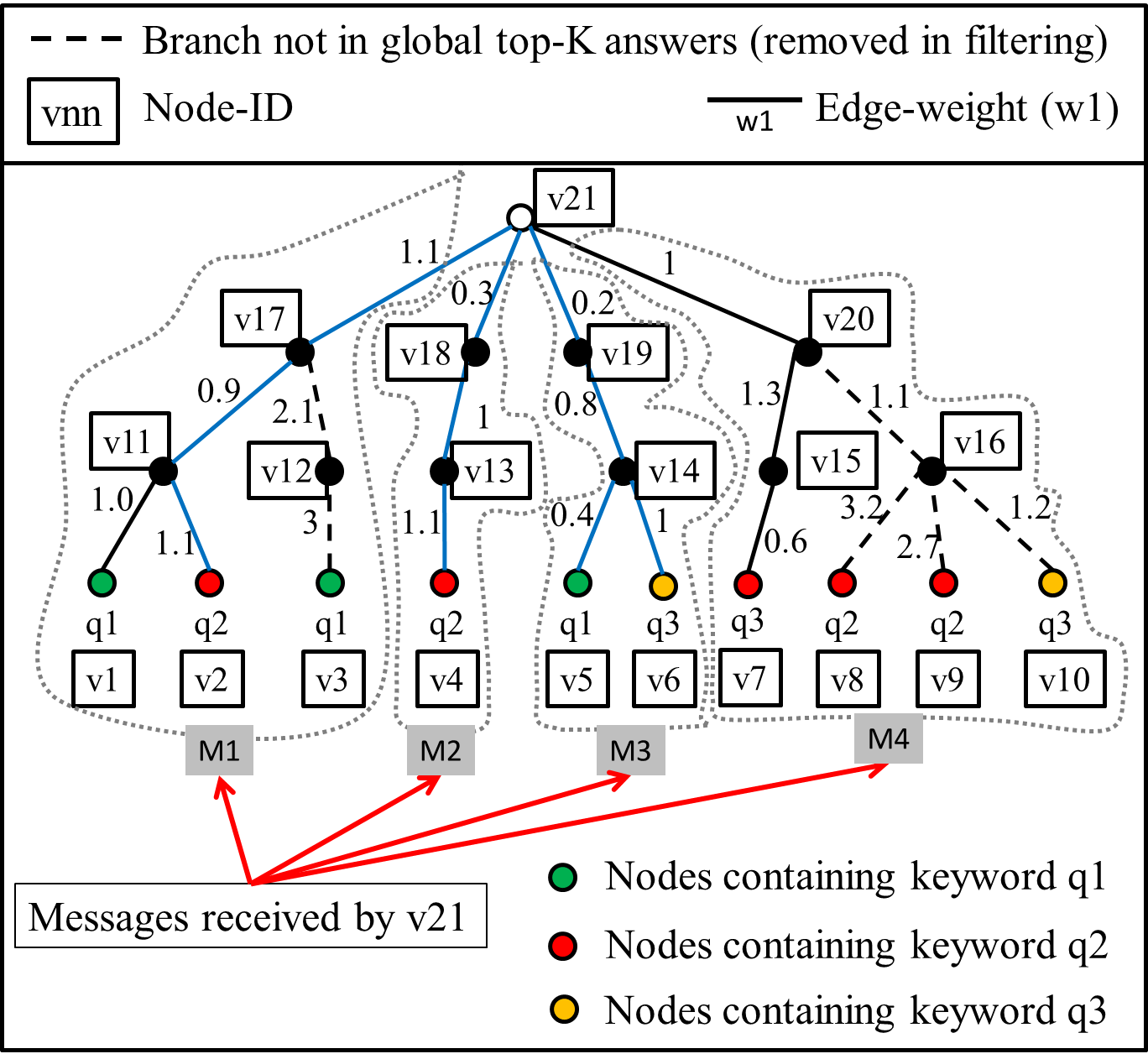}
   \caption{Node $v21$ receives messages $\{M1, M2, M3, M4\}$ from nodes $\{v17, v18, v19, v20\}$ respectively; Resulting local-tree of $v21$ is shown.}
   \label{fig:localTree}
\end{wrapfigure}

\textit{\textbf{Step-1- Receive Messages}}: In a superstep, nodes that receive message(s) become active and all the other nodes remain dormant. The set of paths (in the form of a tree), received as incoming messages, form a tree with the current node at the root. This tree is referred to as \textit{local-tree} of a node, sample local-tree of a node $v_{21}$ is shown in Figure \ref{fig:localTree}.

In the local-tree of a node, there can be more than one subtrees which contain all keywords of the query, e.g., ($\{\{v1, v2, \\
v6\}, \{v1, v2, v10\}, ...\}$) in Figure \ref{fig:localTree}. We drop those branches of a local-tree that are not part of top-K partial answers of any keyword-set in that local-tree, and the remaining tree is called \textit{filtered local-tree}. Such branches are shown by dotted lines in Figure \ref{fig:localTree}, e.g., $v17 - v12 - v3$. If the top-K partial answers of all keyword-sets are retained, we don't miss the global top-K answer as shown in analytical proof in Section \ref{sec:splRetain}. At every node we maintain two data-structures $S_K$ and $V_K$, which contain top-K path-lengths of all keyword-sets $k_i \in \mathcal{P}(Q)$ and the set of node-ids contained in the corresponding trees, respectively. Calculation of the sets $S_K$ and $V_K$ is one of the most compute intensive task of DKS algorithm, since it iterates on the power-set of set of input keywords $Q$, and therefore an optimized approach for calculation of these sets is given in Section \ref{sec:opti}.

\textbf{\textit{Step-2- Evaluate}}: Based on the local-tree of a node it is possible to determine whether the node has a path to keyword-nodes of all the keywords, or not. If yes, it declares itself as the root-node of an answer. We extract local top-K answer-trees from the filtered local-tree of a node, following an approach described in Section \ref{sec:opti}. Next, we describe how to identify the global top-K answer-trees from many such local top-K answer-trees.

\textbf{\textit{Step-3- Sending Aggregator Messages}}: After extraction of the local top-K answers from the local-tree, we calculate the path-lengths of all keyword-sets $k_i$ in every answer and store them in set $L$. Extracted answers and the corresponding set $L$ is sent to an aggregator $A_A$. We also extract the smallest path-length of all keyword-sets $k_i$ from the set $S_K$ and send it to another aggregator $A_S$. Details of these aggregators are in given in Step-5. 

If we were to traverse entire graph it will lead to too many messages being exchanged and DKS may become extremely slow and may never finish. Therefore we need to stop the BFS traversal as soon as possible. We stop BFS traversal when we are certain that the further traversal of the graph will not lead to a better answer-tree than those found so far, this condition is referred to as exit criterion. The aggregated values of above metrics are used for evaluation of exit criterion in Step-6.

\begin{wrapfigure}[23]{R}{0.6\columnwidth}
   \centering
    \includegraphics[width=0.6\columnwidth]{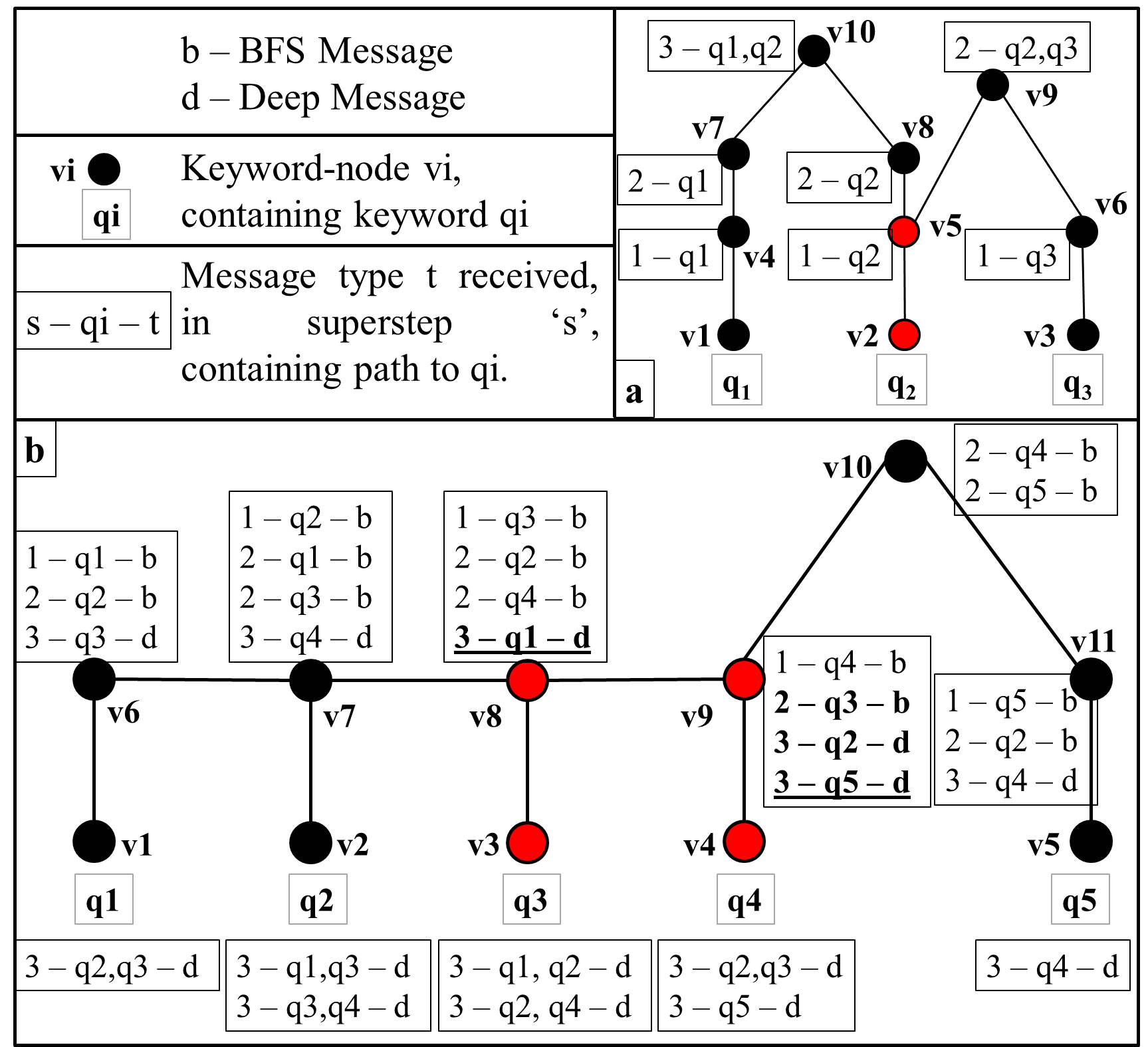}
   \caption{a)~Need for Deep Message, and b)~Need to propagate deep Message}
   \label{fig:needDeep}
\end{wrapfigure}

\textit{\textbf{Step-4- Send BFS / Deep Messages}}: Active nodes of a superstep send filtered local-tree and the sets $S_K$ and $V_K$ of sending node, as a message, to their neighbors for BFS traversal, in order to locate a node which contains paths to all keywords of the query. However, using BFS traversal we can only discover the trees that are balanced at the root-node. For example, in Figure \ref{fig:needDeep}-(a), $v5$ is the root-node and the answer-tree shown here is not balanced at $v5$. When following BFS traversal, $v10$ receives a message from $v7$ and $v8$, but $v7$ remains unaware of the path to $v8$ primarily because in a distributed setting every node is being processed independently on potentially a different worker. Therefore, we send message to both sides of the nodes containing a path to each-other, for example from $v10$ we send a message to $\{v1, v4, v7\}$ containing a path to node $v2$. Similar messages are sent from $v9$ also. As a result, nodes $v5$ and $v2$ get identified as a root-node of an answer-tree. Such messages are called \textit{deep-messages}. The deep messages need to be propagated recursively to cover cases such the one shown in Figure \ref{fig:needDeep}-(b). Note that even after the exit criterion is satisfied, we receive, and propagate the deep messages that were sent before exit criterion was satisfied.

% \begin{figure}
%     \includegraphics[width=\columnwidth]{images/Deep-Prop.png}
%    \caption{A)~Need for Deep Message, and B)~Need for Propagation Message}
%   \label{fig:deep}
% \end{figure}

\textit{\textbf{Step-5- Aggregation:}} We use two aggregators $A_A$ and $A_S$, and the aggregated values of these are used for evaluation of exit criterion described in Setp-6. The $A_A$ aggregator: 
i)~removes duplicate answers, 
ii)~identifies global top-K answers, and 
iii)~calculates the largest path-lengths($L_n$) of all keyword-sets among globally top-K answers, by aggregating the $L$ sets from its messages.
Further, at every active node of a superstep $n$, we calculate a set $S=\{s_i^n, \forall k_i \in \mathcal{P}(Q)\}$, comprising of the smallest path-lengths of all keyword-sets and send it to aggregator $A_S$. In $A_S$ aggregator, we determine the smallest value of these path-lengths, and prepare a set $S_n~=~\{\min({s_i^n}),$ $\forall i = \{1, ..., (2^m-1)\} \}$. The set $S_n$ contains the smallest path-lengths of all keyword-sets $k_i$ in $n^{th}$ superstep. 

\newcommand{\MEst}[3]{\hat{#1}_{#2}^{#3}}

\textit{\textbf{Step-6- Check for Exit}}: If we can say that subsequently discovered answer-trees will have weights more than the best found so far, we can stop traversing further. For this, we estimate the smallest path-length of all keyword-sets in next superstep as $\MEst{s}{i}{n+1}=(s_i^n + e_{min})$. Here, $e_{min}$ is the smallest edge weight in the graph, and therefore $\MEst{s}{i}{n+1} \le s_i^{n+1}$. If all these estimated path-lengths $\MEst{s}{i}{n+1}$ are larger than the corresponding path-length in $L_n$, i.e., $\forall \MEst{s}{i}{n+1} \in \MEst{S}{n}{}$, $ \MEst{s}{i}{n+1}> l_i^n$, $\forall l_i \in L_n$, we can say that all subsequent answer-trees will be worse than those found so far. This condition is referred to as the \textit{exit criterion for BFS traversal}.
%================================================
\section{DKS Optimization and Analysis}
\label{sec:opti}
% !TEX root = ../DKS-FullPaper.tex
Identification of local top-K path-lengths for all keyword-sets (i.e., sets $S_K$ and $V_K$) is one of the most compute intensive tasks in DKS algorithm, therefore we first present an optimized approach for such a calculation followed by analysis of computational and communication cost.

\subsection{Optimization for Local Tree Filtering}
\label{sec:ltfOpti}
To understand the problem of calculation of sets $S_K$ and $V_K$, let is consider the brute-force method first, it will involve steps such as: 
a)~traverse the local-tree and store keyword-wise paths from root-node to the keyword-nodes, 
b)~for every keyword-set $k_i$ generate various combinations of these paths,
c)~for each of such combination find the path-length of corresponding $k_i$, which is not equal to the sum of path-lengths of single keywords contained in the keyword-set $k_i$. This is because some of the edges between two paths may be common. We will therefore need to traverse the local-tree, in order to find the overlapping edges and then to calculate the path-length. 
d)~finally, find the local top-K path-lengths (for every keyword-set $k_i$) from various path-lengths of a keyword-set $k_i$.
All this will require traversing the local-tree exponential number of times (in number of keywords $m$) at every active-node, making it a compute intensive task.

As a first step towards optimization of this process, we maintain two data-structures $S_K$ and $V_K$ at every node, as described in Section \ref{sec:overview}. Further, we assume that in the local-tree of a node there are, on an average, $p$ keyword-nodes for every keyword $q_i$. Therefore, if a keyword-set contains $r (=|k_i|)$ keywords we will have to evaluate $(p \times p ... $r-times$)= p^r$ different trees. There will be ${m \choose r}$ keyword-sets that will contain $r$ keywords in them. For evaluation of top-K path-lengths of all such keyword-sets, we will have to evaluate $p^r \times {m \choose r}$ trees. Therefore, total number of trees that we will need to evaluate for top-K path-lengths of all keyword-sets, in order to fill data-structures $S_K$ and $V_K$ are given in (\ref{eq:treeCount}):

\begin{equation}
 p\times{m \choose 1} + p^2\times{m \choose 2} + ... + p^m 
\times {m \choose m} = (1+p)^m
 \label{eq:treeCount}
\end{equation}

Further, since $p$ can be high, it will become hard to evaluate the data-structures $S_K$ and $V_K$. However, starting from keyword-nodes, if every node maintains these data-structures and also passes these to their neighboring nodes, in the message payload, then there will be two benefits: i)~we will not need to traverse the local-tree to find the answer-trees for various keyword combinations $k_i$, i.e., Steps (a)-(c) of brute-force approach is not required anymore; ii)~the maximum value of $p$ will be ($|\mathcal{M}|\times K$), because each message can contain at the most $K$ keyword-nodes for every keyword $q_i$ and $|\mathcal{M}|$ is the number of messages a node receives. Therefore, total number of path-lengths to be evaluated will be $(1+|\mathcal{M}|\times K)^m$, using Eq \ref{eq:treeCount}.

If we process each message separately, the total number of trees evaluated for a pair of messages will be $(1+ 2K)^m$, and therefore for 
processing all incoming messages at a node we will have to evaluate $|\mathcal{M}|\times (1+2\times K)^m$ trees, which is lesser than 
$(1+|\mathcal{M}|\times K)^m$. Effectively, the time-complexity of preparing the data-structures $S_K$ and $V_K$ at a node will be 
$O\big(|\mathcal{M}|\times (1+2\times K)^m\big)$. Further key benefit of these data-structures is that we can purge the nodes that are 
not present in $V_K$ to obtain \textit{filtered local-tree}. 

\subsection{Time Complexity}
\label{sec:tc}
Computationally, there are two most compute intensive parts of DKS. First is calculation of $S_K$ and $V_K$ at active nodes; secondly for many miscellaneous tasks we need to traverse the local-tree of a node, e.g., purging of extra branches, extraction of top-K answer-trees etc. The \textit{worst-case} time-complexity of the first task, was analyzed in Section \ref{sec:ltfOpti}. Assuming that early exit is not effective and we need to perform this task at every node of the graph, the time-complexity of this task for entire graph will be $O\big(|V| \times \overline{d}\times (1+2\times K)^m\big)$. Here, $|V|$ is the number of nodes in the  graph, and $\overline{d}$ is the average degree of a node in the graph assumed to be equivalent to average number of messages on every node. For second task we need to estimate the average size of the filtered local-tree of a node. For this, assuming every node has small number of child nodes $c$, and height of this tree is $h$, there will be $c^h$ nodes in the local-tree. Such a tree needs to be maintained and traversed at every node of the graph, therefore, the total time complexity of the DKS algorithm can be taken as $O\bigg(\big(|V| \times \overline{d} \times (1+2\times K)^m\big) + \big(c^h \times |V| \big) \bigg)$. It is observed that most of the time $c$ and $h$ are very small integers, i.e.,  $< 5$, and often in keyword searches the number of keywords in a typical search query would be small, therefore the problem becomes tractable. Further, it is evident that the worst case time-complexity of DKS algorithm is linear in the number of nodes in the graph and the number of edges in the graph, while exponential in the number of query keywords. Therefore, if the number of keywords are high or we are interested in too many answers (high value of $K$), DKS algorithm will not perform efficiently.

\subsection{Dist. processing \& Communication Cost}
\label{sec:cc}
If we were to execute the DKS algorithm  in standalone mode, without any fundamental modification, we will run a loop for every superstep. At every frontier node we will combine the filtered local-trees of its neighboring nodes to get its local-tree, instead of getting them as a message as done in distributed implementation. We then filter this tree to get \textit{filtered local-tree}. This will involve the process of calculating the set $S_K$, as performed in the distributed version. Therefore, the primary difference between standalone mode and distributed mode will be that of communication overhead.

To estimate the communication cost, we assume that on an average the number of messages sent by a node are directly proportional to the degree of a node, i.e., linear in the average degree of nodes of the graph, at every node. Therefore total number of messages passed will be directly proportional to $|E|$ the total number of edges in the graph. We discussed the average size of the local-tree of a node to be $c^h$. Therefore, we assume that the total communication cost of DKS algorithm is $(|E| \times c^h)$. Here, $c$ and $h$ are not high since we are not interested in the finding answer-trees with large height.

\subsection{Practical Issues}
\label{sec:practIssues}
It was observed that the system hangs if the total number of messages to be received, in a single superstep, are more than $\sim $ a million (especially after first two supersteps). We stopped subsequent supersteps when this limit was reached and estimated smallest possible answer weight which can get discovered by further exploration of the graph. Ratio of this \textit{estimated smallest possible answer weight} (given below) and the best answer-weight found by our algorithm is reported as \textit{SPA-Ratio}. 

\textbf{\textit{Estimation of smallest possible answer weight}}: At the end of a superstep $S_n$, the set of smallest path-lengths of all keyword-sets, is known and we can estimate $S_{n+1}$ the smallest path-lengths of all keyword-sets in next superstep. From the smallest path-lengths of all keyword-sets we want to estimate the smallest possible answer weight. To construct an answer from all keyword-sets we choose a subset of keyword-sets in such a manner that the chosen keyword-sets collectively contain all keyword sets. We use dynamic programming to exhaustively search the entire search space and find the smallest possible answer weight that can get discovered by further BFS exploration of the graph. We report this answer-weight as the smallest possible answer-weight. The smallest possible answer-weight helps us in estimation of degree of approximation of the detected answer-tree, referred to as \textit{SPA-ratio}, as described above.

%================================================
\section{Analytical Proof}
\label{sec:proof}
In this section we state a theorem about optimality of the answers discovered by our algorithm, and present an analytical proof subsequently. 

\newtheorem{theorem}{Theorem}
\begin{theorem}
\label{thm:dks}
The breadth-first-search traversal on a graph $G~=~(V,E)$ having $w(e) > 0, \forall e \in E$, executed to find top-K minimum steiner trees, can be stopped after $n^{th}$ iteration (superstep), without missing the optimal answer when Eq. \ref{eq:exit} is satisfied. 
\end{theorem}

\newcommand{\Est}[3]{\hat{#1}_{#2}^{#3}}

\begin{equation}
\label{eq:exit}
\Est{s}{i}{{n+1}} > l_i^n; \forall \Est{s}{i}{{n+1}} \in \Est{S}{n+1}{} \text{, } \forall~l_i^n \in L_n  
\end{equation}

Here, $L_n$ is a set of the largest path-lengths of all keyword-sets among the global top-K answers found at the end of $n^{th}$ supserstep; and $\Est{S}{n+1}{}$ is a set of the estimated shortest path-lengths of all keyword-sets for supserstep $(n+1)$, such that $\Est{s}{i}{{n}} \geq s_i^n$, $\forall \Est{s}{i}{{n}} \in \Est{S}{n+1}{}$ \& $\forall s_i^n \in S_{n}$, i.e., the estimated shortest path-length of all keyword-sets for a superstep $n$ should not be less than the corresponding actual shortest path-length of that superstep.

\subsection{Overview and Intuition}
\label{sec:thm}
When searching for the answer-trees, following the algorithm described in Section \ref{sec:overview}, and after we discover first $K$ 
answer-trees at the aggregator, we are not sure whether these are globally optimal. We can stop BFS exploration only when we are sure that further exploration
will not lead to any better answer-tree. For this we need to estimate the smallest possible weight of an answer-tree that can get discovered 
by further iterations of BFS. If this estimated answer weight is more than the largest answer weight found so far then, we need not perform
BFS exploration any further. This evaluation should happen between every two consecutive supersteps.

It is computationally hard to estimate the smallest answer-weight of a subsequent superstep, even if we can estimate the smallest path-lengths of all keyword-sets for that superstep. This is because the keyword-sets are the elements of the power-set of keywords, and many different combination of these keyword-sets can make an answer-tree. Also note that the union of all keyword-sets is the set of all keywords. Therefore, we establish the exit criterion based on Fagin's algorithm\cite{p:faTopK}, brief summary of Fagin's algorithm is given in Appendix \ref{sec:fagin}. To make Fagin's algorithm applicable in our setting, we represent the answer weight as an aggregate function of path-lengths of all keyword-sets, i.e., $w(A) = f(k_1, k_2, ..., k_{2^m-1})$, which increases monotonically with increase in path-lengths. Further, a sorted list of the arguments of this function, i.e., keyword-sets $k_i$ should be present, which is actually not available.

We observed that the shortest path-lengths of all keyword-sets increase monotonically across consecutive supersteps of breadth-first-search, and can work as a proxy for the sorted list. Therefore, we start with Lemma \ref{lemm:spl}, where we prove this formally. As a result, Fagin's algorithm becomes applicable, and therefore in Lemma \ref{lemm:fagin} we state that the answer-tree can be found at a specific set of nodes only, referred to as candidates-nodes. Here, nodes for which $\exists k_i \text{ s.t., }\Est{s}{i}{{n+1}} < l_i^n$, are considered as candidate nodes, i.e., nodes that have the shortest estimated path-length of any keyword-sets in next superstep, smaller than the largest path-lengths of corresponding keyword-set in top-K answer-trees found so far. 

We further restrict the search of candidate nodes with the help of Lemma \ref{lemm:cand}, where we state that it is sufficient to evaluation only those candidate nodes that are at the frontier of the breadth-first-search traversal, which results in better efficiency. Therefore, we keep performing the BFS exploration until no more candidate nodes are left, i.e., Eq. \ref{eq:exit} is satisfied. Further, in Lemma \ref{lemm:localTree} we state that by purging those branches of a local-trees that are not in local top-K, we don't miss any globally optimal answer-tree. As a result, finally aforementioned theorem is proven that the problem of keyword search can be solved using BFS exploration and that we don't need to traverse the entire graph for finding the top-K answer-trees.

\subsection{Proof Details}

\subsubsection{Shortest Path-Length Increases}
\label{sec:spl}

% \newdef{lemma}{Lemma}
\begin{lemm}
\label{lemm:spl}
The shortest path-length of a keyword-set $k_i$ among all frontier-nodes of a superstep, will definitely increase in a subsequent superstep, i.e., $s_i^n \leq s_i^{n+1}$.
\end{lemm}

\begin{wrapfigure}[15]{R}{0.5\columnwidth}
   \begin{tabular}{|l|p{3.6cm}|}
   \hline
   \raisebox{-1\height}{\includegraphics[width=.23\columnwidth]{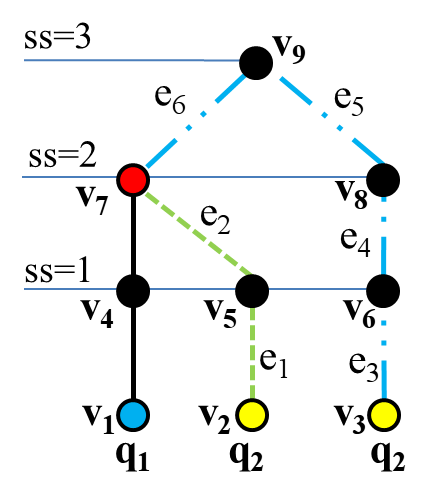}} & \[p_i = w(e_1 + e_2)\] \[p_j = w(e_3 + e_4 + e_5+e_6)\] \[ p_i \leq w(e_3 + e_4)\] \[\implies p_i < p_j\] Here, $w(e_i+e_j)$ is sum of lengths of edges $e_i$ and $e_j$. \\
   \hline
   \end{tabular}   
   \caption{For proof of Lemma \ref{lemm:spl}}
   \label{fig:l1}
\end{wrapfigure}

Here, $s_i^n$ is the shortest path-length of a keyword-set $k_i$, at the frontier-nodes of superstep $n$. 
As explained below via Figure \ref{fig:l1}, during BFS traversal at any node($v_7$) two different types of paths to a keyword-set can get discovered, based on two different types of messages received, i.e., BFS message and deep message. We prove the above Lemma for both of these cases, for $w(e)>0, e \in E$.

\noindent\textbf{\textit{Proof Case (i)}}:  For the first types of paths to keyword-sets, e.g., $v_2 \rightarrow v_5 \rightarrow v_{7}$, it is straight forward to understand that the shortest path-length of any keyword-set can only increase in subsequent a superstep, i.e., path-length of $k_i=\{q_2\}$ at $v_9$ will definitely be more than that at $v_7$. 

\noindent\textbf{\textit{Proof Case (ii)}}: For paths received through deep messages, e.g., $v_3 \rightarrow v_6 \rightarrow v_8 \rightarrow v_9 \rightarrow v_7$, we want to prove that: If a node $v$ has shortest path-length ($p_i$) for a keyword-set $k_i$ in superstep $s$, and we get to know of another path to the same keyword-set $k_i$ at $v$ through deep-traversal in any subsequent superstep ($s+\Delta s$), the path-length ($p_j$) of this new path will be more than $p_i$, i.e., $p_i < p_j$.

We can state that $p_j \geq (p_i + e_{min}*\Delta s)$. Here, $e_{min}$ is the smallest edge weight in the graph, and $\Delta s$ is the difference between the superstep numbers in which of $p_i$ and $p_j$ were discovered. This can be asserted because in superstep $s$, a part of this new path would have been discovered. In our example, path $v_4 \rightarrow v_8 \rightarrow v_{12}$ was discovered in $2^{nd}$ superstep itself. The path-length of that part of the path would be more than or equal to the shortest path $p_i$. Since both $e_{min}$ and $\Delta s$ are positive. Therefore the shortest path-length of any $k_i$ increases by at-least $e_{min}$ in every superstep.

\textit{Note:} Using a similar argument, we can also state that the shortest path-length of a keyword-set $k_i$, among all actives nodes of supserstep, can occur only at the frontier-nodes of a superstep. 

\subsubsection{Identify Candidate Nodes}
\begin{wrapfigure}[20]{R}{0.6\columnwidth}
   \centering
   \includegraphics[width=0.6\columnwidth]{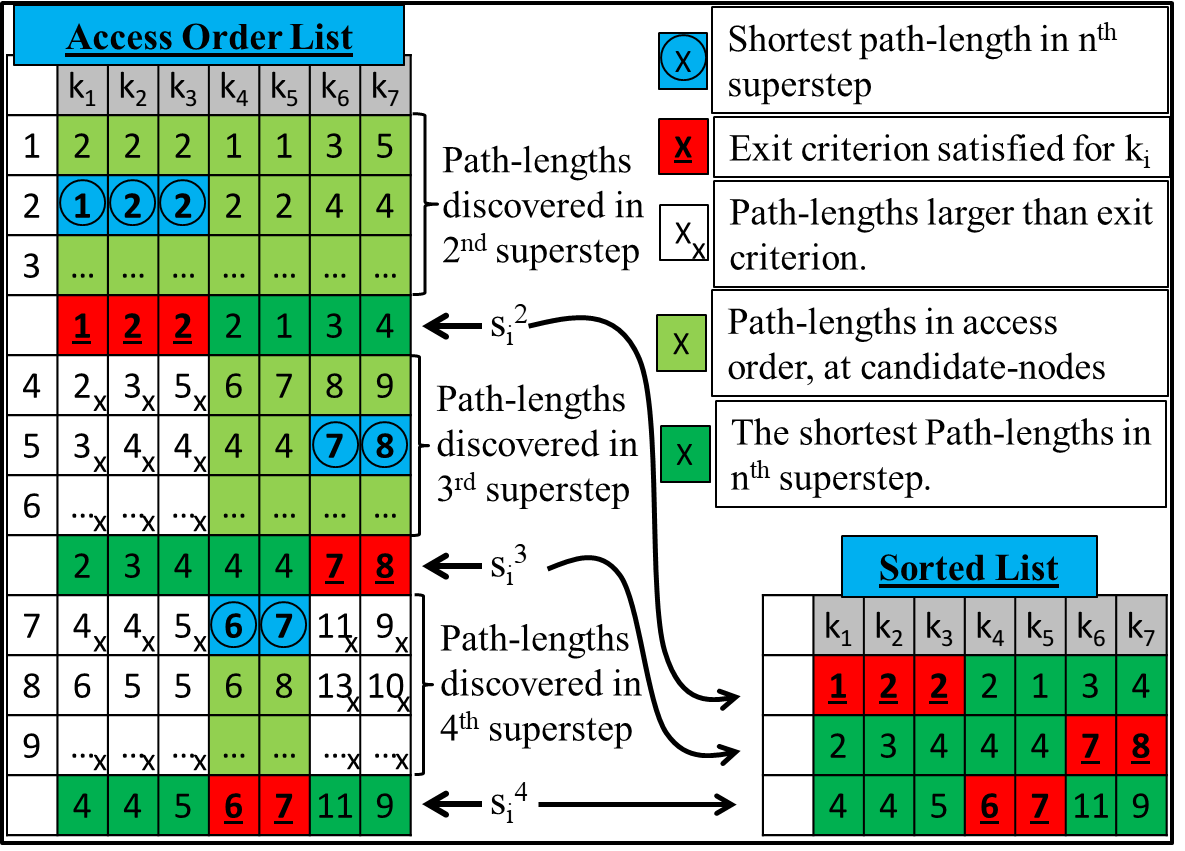}
   \caption{How a sorted list of the shortest path-lengths is extracted from the path-lengths available in access order.}
   \label{fig:sortedList}
\end{wrapfigure}

An overview of Fagin's algorithm \cite{p:faTopK} is given in Appendix \ref{sec:fagin}, which forms the basis for the next Lemma. To apply Fagin's algorithm in this setting, we need to have sorted lists of the input arguments of the aggregate function $w(A)$, and the aggregate function should be monotonic w.r.t. its input arguments. Since the shortest path-length of all keyword-sets increases in every subsequent superstep, we can imagine that for every keyword-set $k_i$, a sorted list exists comprising of the shortest path-lengths of all keyword-sets for every superstep, as explained in Figure \ref{fig:sortedList}. We also define weight of an answer-tree $A$ as a function of path-length of all keyword-sets at the root-node of the answer-tree, as given in Eq. \ref{eq:ansWt}. As a result, Fagin's algorithm becomes applicable, we can identify candidate-nodes, and subsequently establish exit criterion.

\begin{equation}
 \label{eq:ansWt}
w(A) = f(k_1, ..., k_{2^m-1}) =\sum_{\substack{k_i \in \mathcal{P}(Q)}} I(k_i) \times \phi(k_i,A)
\end{equation}
Where, \[ I(k_i) = \left\{
  \begin{array}{l l}
    1 & \quad \text{if, $k_i$ is a constituent-keyword-set}\\
    0 & \quad \text{otherwise}
  \end{array} \right.\]

Here, \textit{constituent-keyword-set} is a keyword-set comprising of all keywords of a sub-tree rooted at child-nodes of the root-node of the local-tree.  The function $w(A)$ given in Eq. \ref{eq:ansWt} can be proved to be monotonic, because it is a conditional summation of strictly positive input arguments. Next, we define a set of \textit{candidate-nodes} $C_n$, based on Fagin's algorithm; it is a set of nodes that can be part of the global top-K answers-trees, after $n$ supersteps. 

\begin{lemm}
\label{lemm:fagin}
After any $K$ answers are found at the aggregator, the set of \textit{candidate-nodes} $(=\mathcal{C})$ comprises of the nodes for which $\exists k_i \text{ s.t., }\Est{s}{i}{{n+1}} < l_i^n$, i.e., at-least one of the estimated shortest path-lengths is smaller than corresponding path-length from the set $L_n$.
\end{lemm}

For example, in Figure \ref{fig:sortedList}, cells marked with a circle are the path-lengths of the set $L_n$, and path-lengths at candidate-nodes are marked in light-green color. Here, it is important to note that after finding K answers-trees in a superstep, we can find a better 
answer in a subsequent superstep, since we explore nodes based on BFS and not in the increasing order of their constituent path-lengths.

\subsubsection{BFS Stopping Criterion}
According to Fagin's algorithm remaining attributes of the candidate objects should be accessed in random order, to identify the global top-K answer-trees.
However, in our setting the scenario is different from that described in Appendix \ref{sec:fagin}, in a manner that there can be more than one value of the same attribute (path-length of a $k_i$) at any node. We need to consider all such path-lengths that are smaller than the corresponding element in set $L_n$. Therefore, instead of performing the random access of remaining path-lengths, we continue to perform BFS exploration until the shortest path-lengths of all keyword-sets at \textit{frontier-nodes} are larger than or equal to the corresponding elements from the set $L_n$. The nodes in the candidate set get revised in every subsequent superstep, to this effect we present next Lemma.

\begin{lemm}
\label{lemm:cand}
Random access of remaining attributes of \\candidate-nodes, is equivalent to random access of remaining attributes at 
frontier-nodes of subsequent supersteps or \textit{traversed candidate nodes} that receive a message.
\end{lemm}

Here, the candidate nodes that were traversed at-least one superstep before the current one, are referred to as \textit{traversed candidate nodes}. 
We prove this Lemma with the respect to two types of nodes: (i)~frontier nodes of previous superstep and (ii)~traversed candidate-nodes. 

\textit{Frontier nodes of previous superstep}: At-least one of the neighboring node of a candidate-node also needs to be part of an answer-tree for the candidate-node to be part of the answer-tree. Since the frontier-nodes of current superstep are neighboring nodes of the frontier-nodes of previous superstep, Lemma \ref{lemm:cand} is proven for frontier nodes of previous superstep.

\begin{figure}
\centering
   \begin{tabular}{|l|p{4.5cm}|}
   \hline
   \raisebox{-1\height}{\includegraphics[width=.3\columnwidth]{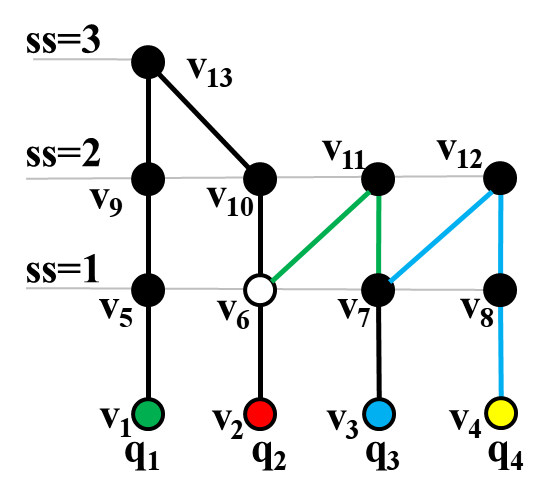}} & \begin{itemize}
                                                                                      \item Path to $q_1$ reaches $v_6$ in \textit{ss=4}, via deep message.
                                                                                      \item Path to $q_4$ reaches $v_7$ in \textit{ss=3}, and it reaches $v_6$ in \textit{ss=4}, via deep-messages.                                                                                      
                                                                                     \end{itemize} \\
   \hline
   \end{tabular}   
   \caption{For proof of Lemma \ref{lemm:cand}}
   \label{fig:l3}
\end{figure}

The \textit{traversed candidate nodes} can become root-node of an answer-tree with the help of two types of paths, first when the new path passes through a frontier-node and the second when the new path does not pass through one of the frontier-nodes. For example, as shown in Figure \ref{fig:l3}, let us consider a \textit{traversed candidate node} $v_6$ in $4^{th}$ superstep. Here, path to $q_1$ at $v_6$ passes through frontier-node $v_{13}$, and path to $q_4$ at $v_6$ does not pass through any frontier-node. For the first types of path no further proof is needed. The second types of paths, will get taken care by deep messages at \textit{candidate traversed nodes}. Therefore, Lemma \ref{lemm:cand} is proved, and we continue to BFS exploration of the graph until none of the frontier-nodes is a candidate-node, and stop DKS algorithm when no more deep-messages are left to be passed around.
\subsubsection{Global Vs Local TOP-K Steiner Tree}
\label{sec:splRetain}
\begin{wrapfigure}[17]{R}{0.6\columnwidth}
   \centering
   \includegraphics[width=0.6\columnwidth]{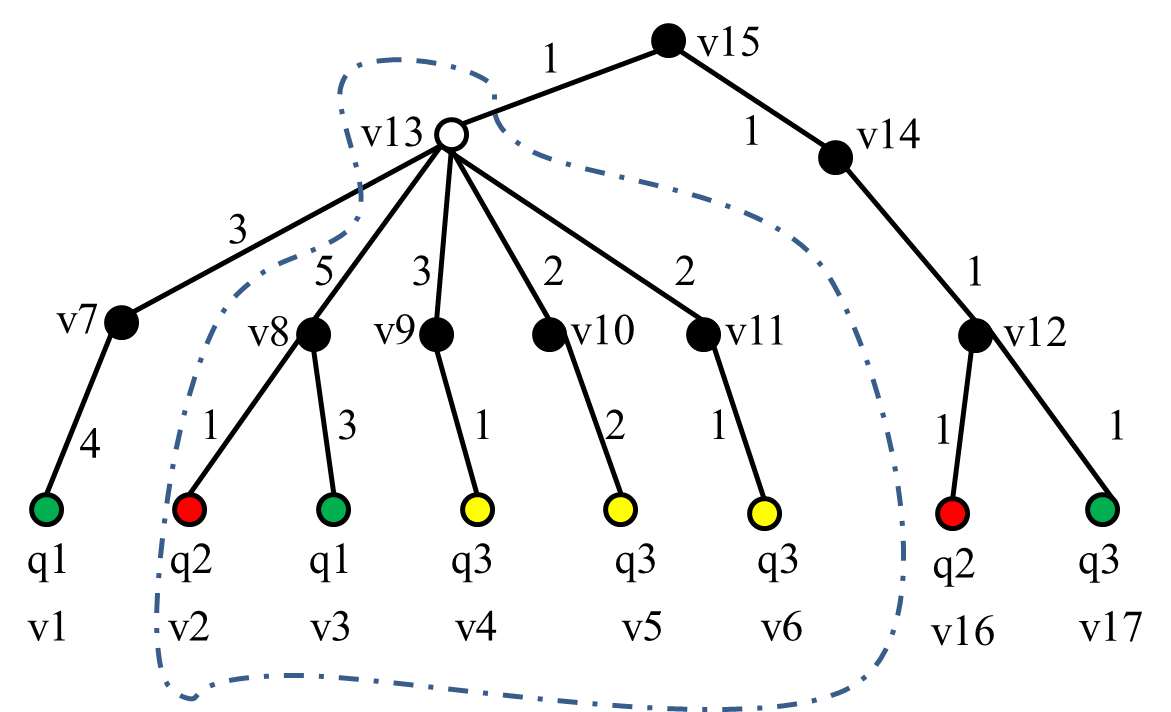}
   \caption{shortest path to $q1$ not in top-K answers}
   \label{fig:branchFiltering}
\end{wrapfigure}

In this section, we analyze the effect of filtering the unwanted branches of the local-tree on the process of finding the minimum steiner tree. This analysis also presents a basis for calculation of set $S_K$ at every node and also for not rejecting some of the branches of the local tree, even if they are not part of any of the local top-K answers. 

\begin{lemm}
\label{lemm:localTree}
 By purging the extra branches of a local-tree, i.e., branches that are not part of local top-K trees of any keyword-set $k_i \in 
\mathcal{P}(Q)$, we don't miss any of the top-K answer-tree.
\end{lemm}

Here, it is important to note that if the branches of a local-tree that are not in local top-K answer-trees are purged we can miss an answer-tree. This can be observed from the example shown in Figure \ref{fig:branchFiltering}. Here for $K=3$, the branch $\{v13, v7, v1\}$ is not in top-3 answer-trees at node $v_{13}$  but if not purged at $v_{13}$, it can be part of a global top-K answers rooted at vertex $v_{15}$. Therefore branches of a local-tree that are part of the top-K partial answer-tree of any of the keyword-set $k_i \in \mathcal{P}(Q)$ should not be purged. Further, it is trivial to prove the remaining argument of this Lemma, that by purging all the remaining branches of a local-tree we don't miss any of the top-K answers.

In summary, we have proven that the answer weight can be represented as a monotonic function of path-lengths of all keyword-sets of a keyword-query, and that BFS way of searching the graph for answer-trees, is equivalent to sorted access of path-lengths w.r.t. shortest path-lengths at frontier-nodes in consecutive supersteps. Therefore, Fagin's algorithm becomes applicable in this setting and as stated by Fagin, we can stop this exploration without missing the optimal answer-tree. We also presented proof for Lemmas \ref{lemm:spl}, \ref{lemm:fagin}, \ref{lemm:cand}, and \ref{lemm:localTree} and therefore Theorem \ref{thm:dks} is proved.
%==========================Distributed Keyword Search using Pregel======================
\section{Experiments and Analysis}
\label{sec:exp}
\subsection{Data, Infrastructure and Implementation}

\begin{wrapfigure}[18]{R}{0.6\columnwidth}
  \includegraphics[width=0.6\columnwidth]{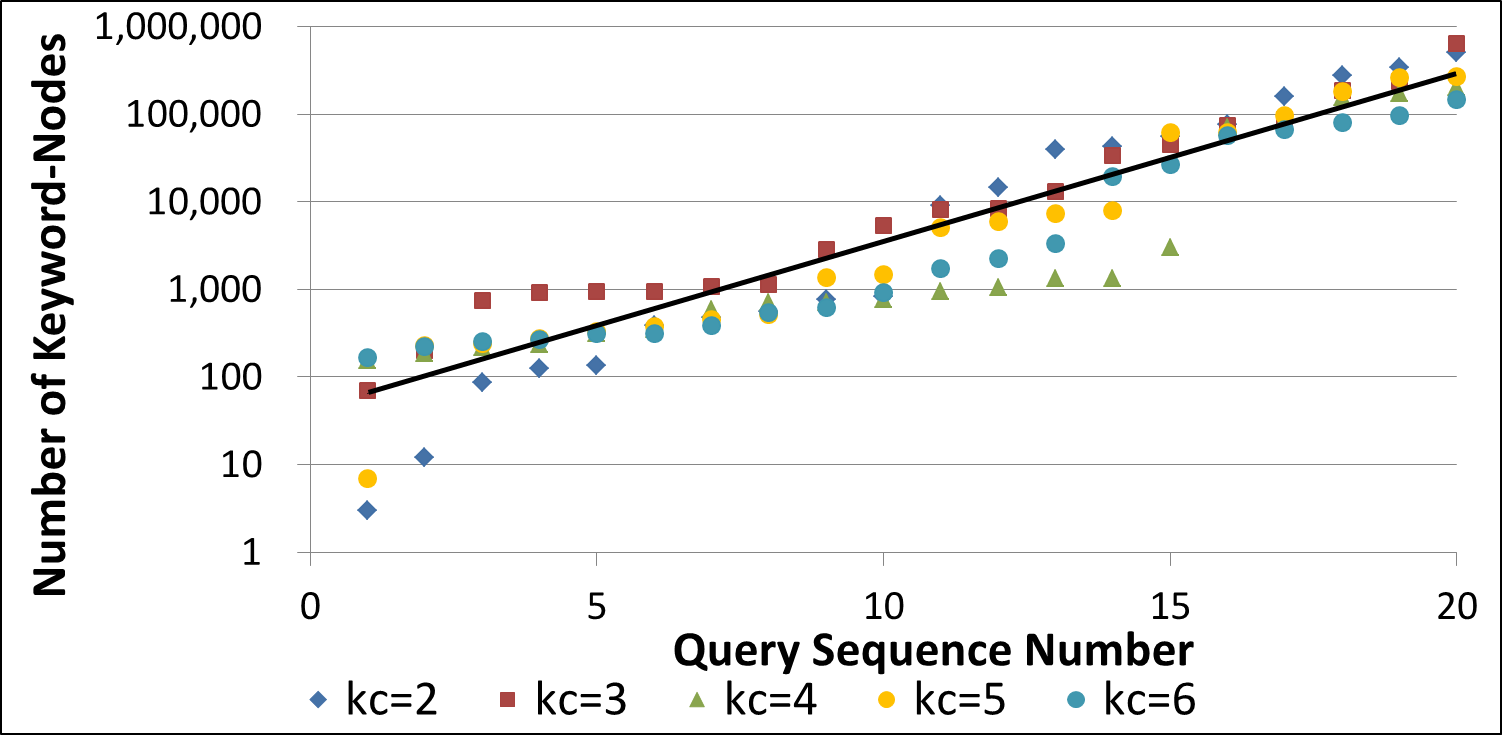}
  \caption{Number of keyword-nodes (on a log-scale) for the 20 queries for various keyword counts. Trendline shown on exponential scale.}
  \label{fig:qr}
 \end{wrapfigure}

\textbf{\textit{Datasets and Infrastructure}}: We performed experiments on two datasets of Linked-Open Data \cite{p:lod}: a)~sec-rdfabout, RDF data about U.S. securities and  corporate ownership (460,451 nodes and 500,384 edges); and b)~bluk-bnb, RDF data on British National Bibliography (16.1 million nodes and 46.6 million edges). Bluk-bnb is not only the largest dataset on which keyword-search has been attempted in the research literature, but also larger than what can run on systems such as BANKS \cite{p:banks1,p:banks3}. Following the strategy proposed by Coffman et al. in \cite{p:JoelCoffmanValidation}, we generated 100 queries for bluk-bnb dataset based on the frequently occurring keywords. These 100 queries were generated such that first 20 queries contained two keywords each, next 20 queries contained three keywords each, and so on, i.e., the number of keywords per query varied from 2 to 6 in these 100 queries. The 100 queries thus obtained were used for running all the experiments reported in this paper. Further, keywords for these queries were chosen in a manner that the total number of keyword-nodes per query, varied from a small number $(\sim 10)$ to a large number $(\sim 500,000)$ as shown in Figure \ref{fig:qr}; here, it can be observed that the number of keyword-nodes increase exponentially across different queries. All experiments reported in this paper were conducted on a compute cluster of four machines, each having 4 Intel Xeon E7520@1.87GHz CPUs with 4 cores, 32 GB RAM, configured to have 35 workers.

\begin{wrapfigure}[36]{R}{0.7\columnwidth}
  \includegraphics[width=0.7\columnwidth]{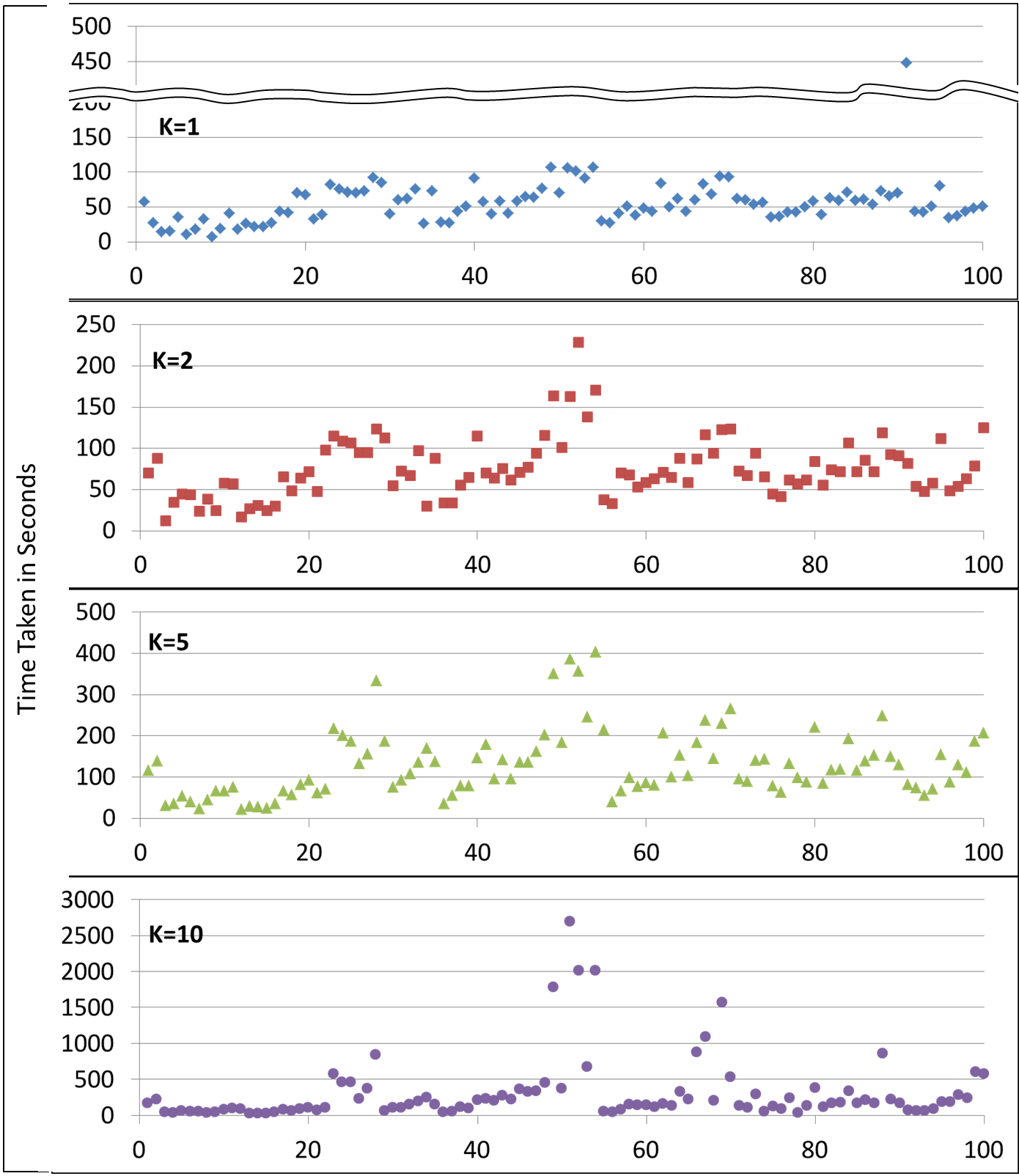}
  \caption{x-axis - Queries, y-axis - Normalized Time taken in seconds for different values of $K=\{1,2,5,10\}$}
  \label{fig:tt}
 \end{wrapfigure}

\textbf{\textit{Implementation}}: We implemented the DKS using an open-source Pregel package Apache Giraph 1.0\footnote{http://giraph.apache.org/} \cite{p:giraph}, which was configured to have 34 workers and a master worker. The edge weights were modeled following a strategy similar to that proposed in \cite{p:banks1}. Here, the edge weight is smaller if the in-degree of the target node is smaller, based on an intuition that if a node $v_1$ has say 10 incoming edges, and another node $v_2$ has 100 incoming edges, the neighboring-nodes of $v_1$ are closer to it as compared to the distance between neighboring-nodes of $v_2$ and $v_2$ itself. In DKS implementation, the edge weights are drawn from a step-function w.r.t. degree of its target node. If the degree $d$ of the target node is smaller than a prior threshold $\tau$ the edge weight is assumed to be \textit{int}($\log_{10} d$) and infinite otherwise. $\tau=1001$, was chosen from the degree distribution of the graph. Also, it was observed that the system hangs if the number of messages in a single superstep are more than approx. 1 million. In such situation, we stopped subsequent supersteps and estimated smallest possible answer weight which can get discovered by further exploration of the graph, following the method presented in Section \ref{sec:practIssues}, above.

\subsection{Benchmarks}
Benchmarks presented in this paper were conducted on the bluk-bnb dataset, while that on sec-rdfabout were presented in a previous paper \cite{p:dks:cikm}. Here, we evaluate i)~the efficiency of our approach by observing the time-taken to run the $100$ queries on bluk-bnb dataset and compare it with the time taken by vanilla parallel BFS implementation, because it is possible to run vanilla parallel BFS on a graph efficiently. Other approaches such as BANKS \cite{p:banks1} are not being compared primarily because those algorithms cannot handle such large volumes of data and our algorithm may not perform well on small datasets. We also present break-up of time taken by various components of DKS described in next paragraph; ii)~Degree of approximation for situations when we had to exit before the exit criterion was satisfied; iii)~the effectiveness of early exit by observing the \% of nodes explored for every query, and for various values of K in top-K; iv)~the communication cost by measuring ratio of total number of messages exchanged and edge-count of the graph; and v)~the effectiveness of the distributed processing by observing the time-taken by a select set of queries by varying the number of worker nodes (compute nodes) in Apache Giraph installation. In all figures of this section, queries are organized in increasing order of keyword-count and keyword-node count.

\textbf{\textit{Time-Taken}}: The vanilla parallel BFS was observed to take approximately 2 min 10 sec. The 90th percentile of the queries take 85 sec ($< 2$ min) for $K=1$, 116 sec ($< 2$ min) for $K=2$, 221 sec ($< 4$ min) for $K=5$, and 609 sec ($\sim 10$ min) for $K=10$ to run the DKS algorithm. Here, time taken for instantiation of worker node jobs in Apache Giraph, first time loading of the graph, and serialization of the final results, i.e., time taken by the system ($120$ sec) has been discounted from the reported running time of DKS algorithm. This normalized running time of the DKS algorithm for all the $100$ queries has been shown in Figure \ref{fig:tt}. Here, it is important to note that the running time of DKS not only depends on the number of keywords, but also on the number of keyword-nodes of the query. However, it can be observed that while the number of keywords increase exponentially but the time-taken does not increase in the same order. 

\begin{table}
\centering
\caption{Percentage of Time taken by DKS components, for different values of $K$}
\begin{tabular}{|c|p{1.7cm}|p{1.7cm}|p{1.7cm}|p{1.7cm}|p{1.5cm}|} \hline
K & Send BFS Msgs & Receive Msgs & Send Deep Msgs & Send Agg Msg & Evaluate\\ \hline \hline
1 & 38\% & 44\% & 6\% & 11\% & 1\%\\ \hline
2 & 37\% & 38\% & 17\% & 8\% & 1\%\\ \hline
5 & 35\% & 37\% & 22\% & 5\% & 1\%\\ \hline
10 & 31\% & 42\% & 21\% & 4\% & 1\% \\ \hline
\hline\end{tabular}
\label{tab:breakup}
\end{table}

\begin{wrapfigure}[15]{R}{0.6\columnwidth}
  \includegraphics[width=0.6\columnwidth]{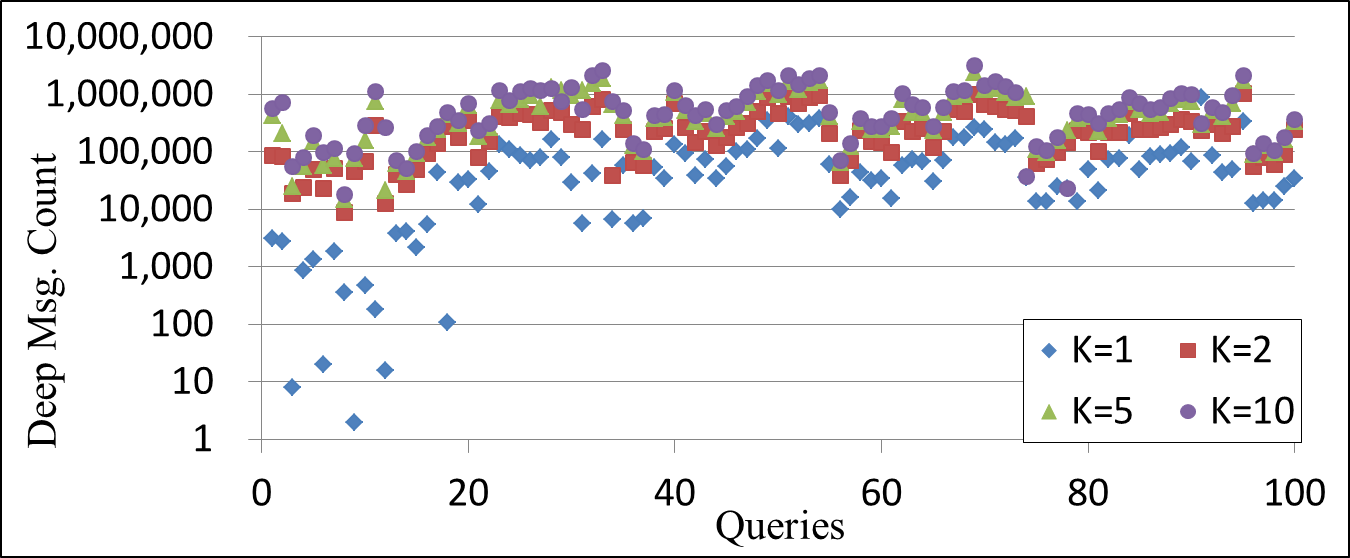}
  \caption{Deep Message Count for all 100 queries, shown varying with increasing values of $K$.}
  \label{fig:dmc}
 \end{wrapfigure}
We also ran various query in performance collection mode only, to measure the time taken for fine-grained steps of the DKS algorithm to understand what part of the algorithm takes most of the time. We divide the DKS algorithm in five components, which are: (i)~Send BFS Message: concerns with iterating over the outgoing edges of a node, serializing the local-tree of the node, and sending the message, (ii)~Receive Message: concerns with Step-1 of DKS, described in Section \ref{sec:overview} which includes the task for calculation of sets $S_K$ and $V_K$ and filtering of local-tree, (iii)~Send Deep Message: this includes iterating over the local-tree, and sending suitable deep-messages (iv)~Extract top-K messages from local-tree of a node, based on set $S_K$ and $V_K$. The results of this analysis have been presented in Table \ref{tab:breakup}. Here, we can observe that most of the time is taken by receive-message step, which is expected based on the analysis presented in Section \ref{sec:opti}. Sending the messages is also a time consuming task because it involves serialization of the local-tree as well as the communication cost. Also with increase in the value of $K$ the time-taken for sending deep-messages increases, primarily because more deep messages were passed as shown in Figure \ref{fig:dmc}.

\begin{figure}
  \centering
  \includegraphics[width=0.6\columnwidth]{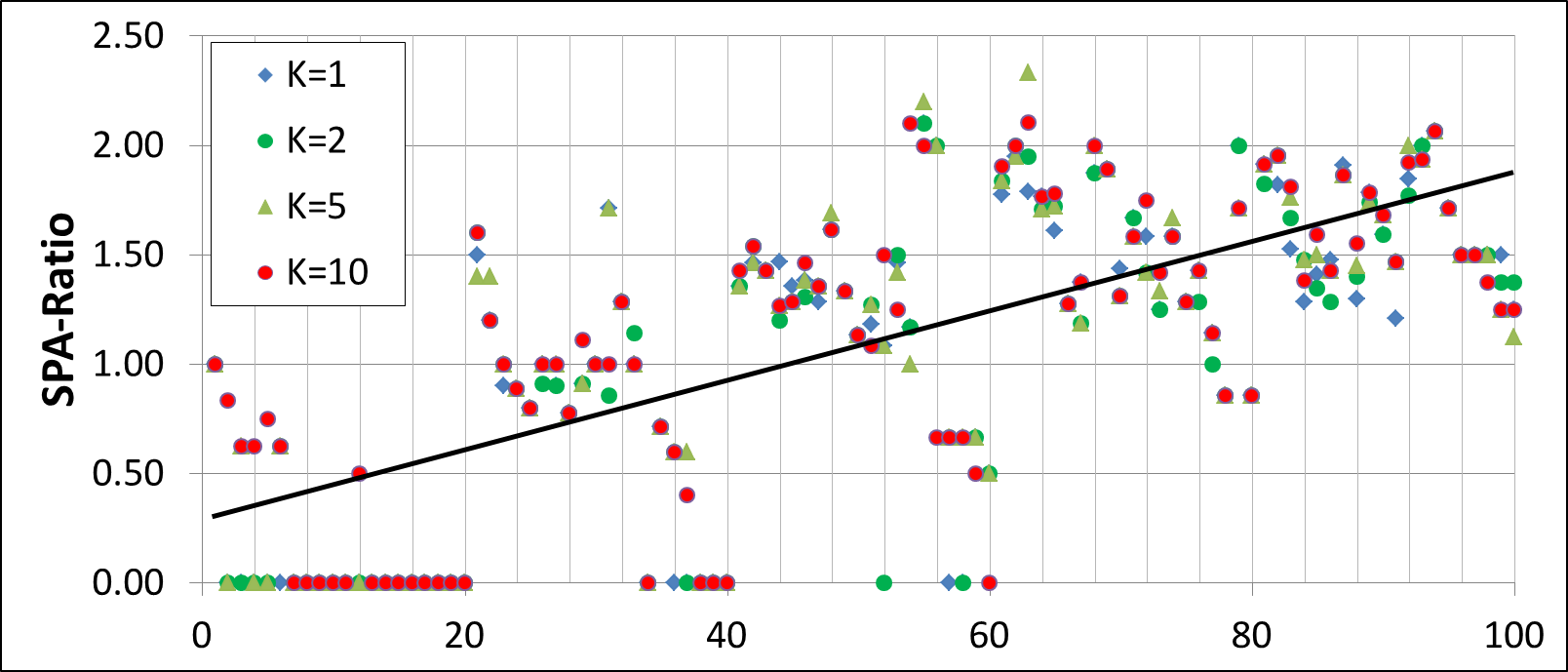}
  \caption{SPA-Ratio shown (on y-axis) for queries organized in increasing order of keyword-count and keyword-node count (on x-axis).}
  \label{fig:spa}
\end{figure}

\textbf{\textit{SPA-Ratio}}: For situations when our Infrastructure was not sufficient enough to tackle the load, we stopped the DKS algorithm after estimating the smallest possible weight that can get discovered by further traversal of the graph. We report the SPA-ratio of the queries, which is a ratio of the weight of the best detected answer-tree to the weight of the smallest possible answer weight that can be detected by further exploration of the graph. Here, for situations where optimal answer was detected the SP-ratio is marked as zero. The SPA-ratio is not the approximation ratio of the algorithm, because deep messages are also stopped, still it can be taken as a measure of the degree of optimality of the detected answer.  The SPA-ratio of all the 100 queries is shown in Figure \ref{fig:spa}, and it was observed that the $90^{th}$ percentile of the SPA-ratio was $\{1.85,1.86,1.89,1.90\}$, for $K=\{1,2,5,10\}$, while the best reported approximation ratio for a heuristic solution is 1.55 \cite{p:SteinerSoln2000} but it has a quadratic time-complexity in the number of nodes of the graph while our approahc is linear in the number of nodes of the graph.

\begin{figure}
  \centering
  \includegraphics[width=0.7\columnwidth]{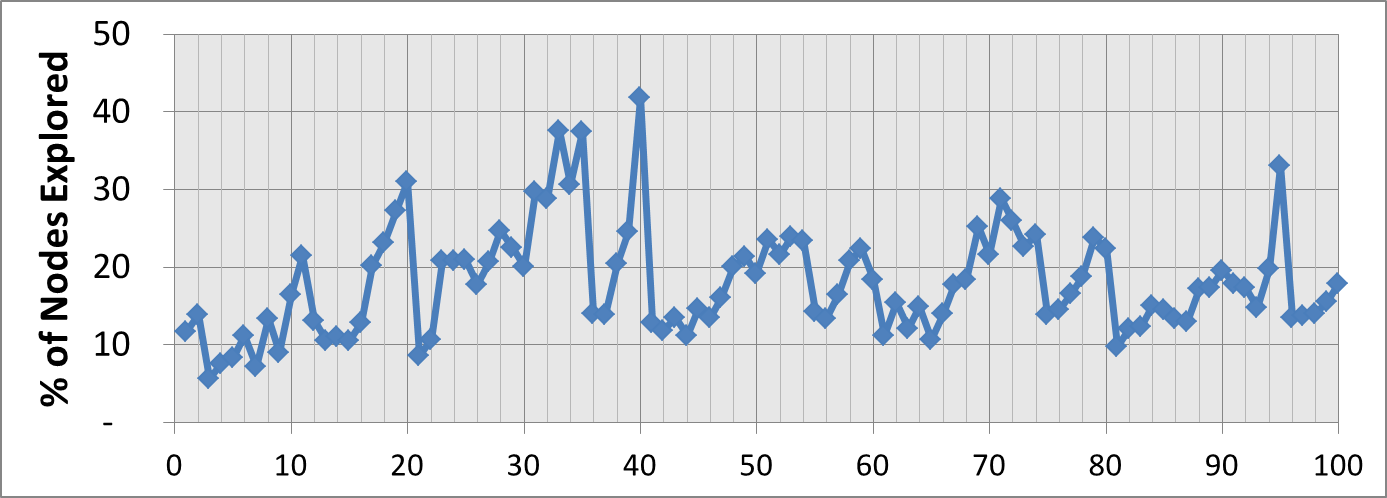}
  \caption{\% of nodes explored w.r.t. number of nodes in the graph, averaged for $K=\{1,2,5,10\}$, for all 100 queries on bluk-bnb dataset. }
  \label{fig:nodeXpl}
 \end{figure}

\begin{figure}
\centering
  \includegraphics[width=0.7\columnwidth]{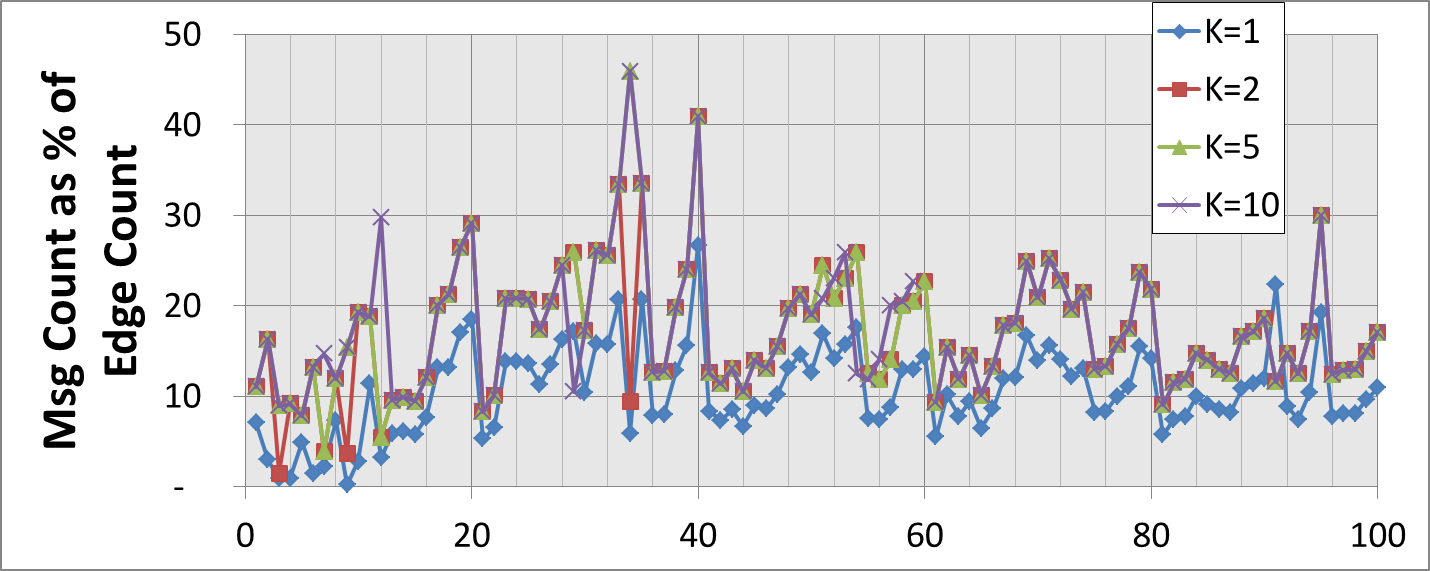}
  \caption{Total number of messages as percentage of $|E|$, shown varying with respect to different values of $K$, for all the 100 queries on bluk-bnb dataset.}
  \label{fig:msgCount}
 \end{figure}

\textbf{\textit{Effectiveness of early exit \& Communication Cost}}: The \% of nodes explored did not show significant change with respect to different values of $K$. Therefore, we present average of the percentage of nodes explored for every query in Figure \ref{fig:nodeXpl}. Here, $90^{th}$ percentile of the percentage of nodes explored was observed to be $\sim 26\%$, indicating that our approach is quite effective in reducing the search space of the problem. In Figure \ref{fig:msgCount}, we have shown the total number of messages exchanged as a \% of the total number of edges in the graph. We observed that the $90^{th}$ percentile of the percentage of message-count with respect to number of edges was $\sim \{16\%, 25\%, 25\%, 26\%\}$ for different values of $K$. This indicates that the number of messages required to be exchanged increase with increase in value of $K$. Further, this experiment corroborates the assumptions made in Section \ref{sec:tc} and \ref{sec:cc} regarding total number of nodes and messages for estimation of time-complexity of our algorithm.

\begin{figure}
\centering
\includegraphics[width=0.7\columnwidth]{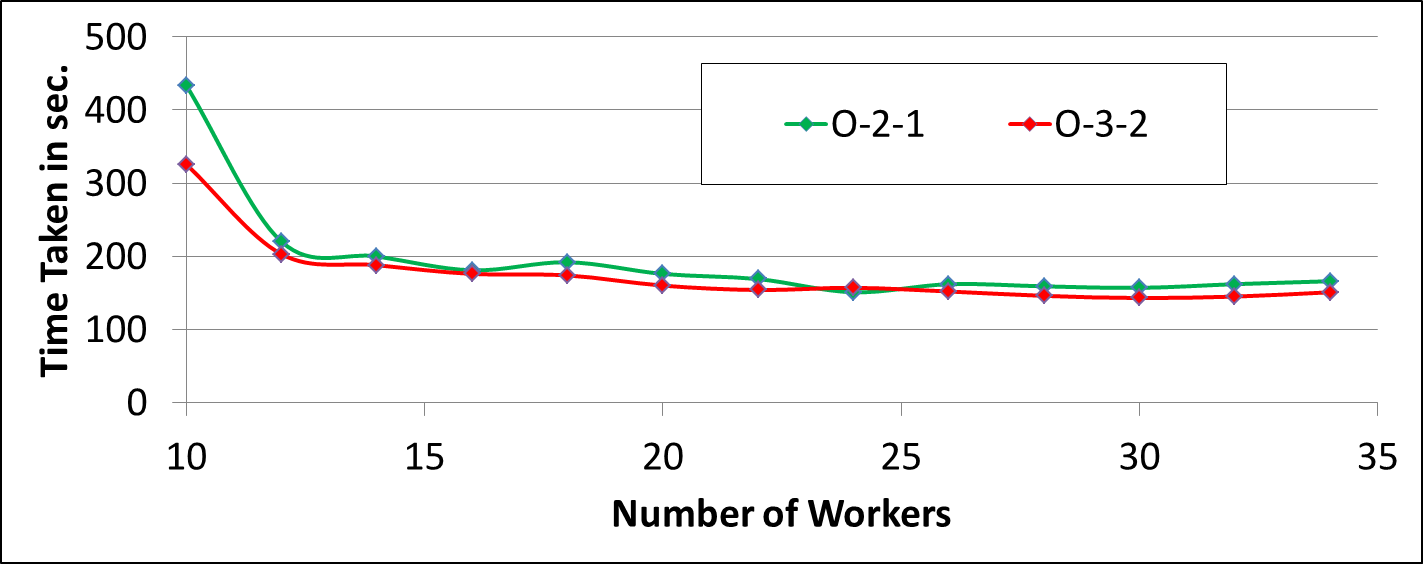}
  \caption{Parallel efficiency of DKS, for two queries both resulting in optimal answers}
  \label{fig:dis}
\end{figure}

\textbf{\textit{Benefits of Distributed Processing}}: Finally we demonstrate the benefits of distributed processing by running same set of queries on different number of worker nodes of the computer cluster. The results of this experiment are shown in Figure \ref{fig:dis}, for a pair of queries with keyword count 2 and 3, respectively. Here, it can be observed that on 3 times the number of worker nodes, the time-taken becomes more than half. However, by increasing the number further not much of gain is observed.
%================================================
\section{Conclusion \& Future Work}
\label{sec:conslusion}
% !TEX root = ../DKS-FullPaper.tex
We have described a novel parallel algorithm for relationship queries on large graphs (equivalent to the group Steiner tree problem). 
Our distributed keyword search (DKS) algorithm is defined in the graph-parallel (Pregel-like) computing paradigm. 
While DKS searches for the root-vertex of an answer-tree following a BFS strategy, the
algorithm ensures that only a fraction of the graph needs to to be explored for most queries. We include analytical proof of optimality
as well as show that even with early exit from BFS we do not miss an optimal answer-tree. We also describe an optimized implementation
of the basic algorithm and analyze its time-complexity. We have also demonstrated that DKS works efficiently on large real-world graphs
derived from linked-open-data, via experimental results on the graph-parallel framework Giraph.

%\end{document}  % This is where a 'short' article might terminate

% ensure same length columns on last page (might need two sub-sequent latex runs)
%\balance

%ACKNOWLEDGMENTS are optional
\section{Acknowledgments}
We thank Prof. Amitabha Bagchi, IIT Delhi for his reviews and suggestions.

% The following two commands are all you need in the
% initial runs of your .tex file to
% produce the bibliography for the citations in your paper.
\bibliographystyle{abbrv}
\bibliography{DKS-Bibtex}

\begin{thebibliography}{10}

\bibitem{p:dks:cikm}
P.~Agarwal, M.~Ramanath, and G.~Shroff.
\newblock Distributed algorithm for relationship queries on large graphs.
\newblock In {\em Proceedings of the workshop on Large-Scale and Distributed
  System for Information Retrieval in CIKM}, LSDS-IR '15.

\bibitem{p:dbXplore}
S.~Agrawal, S.~Chaudhuri, and G.~Das.
\newblock Dbxplorer: enabling keyword search over relational databases.
\newblock In {\em Proceedings of the ACM SIGMOD international conference on
  Management of Data}, SIGMOD '02.

\bibitem{p:giraph}
C.~Avery.
\newblock Giraph: Large-scale graph processing infrastructure on hadoop.
\newblock {\em Proceedings of the Hadoop Summit. Santa Clara'11}.

\bibitem{p:Objectrank1}
A.~Balmin, V.~Hristidis, and Y.~Papakonstantinou.
\newblock Objectrank: Authority-based keyword search in databases.
\newblock In {\em Proceedings of the international conference on Very Large
  Data Bases}, VLDB '04.

\bibitem{p:distSTP1}
F.~Bauer and A.~Varma.
\newblock Distributed algorithms for multicast path setup in data networks.
\newblock {\em IEEE/ACM Transactions on Networking '96}.

\bibitem{p:steinerSurvey2013}
M.~Bezensek and B.~Robic.
\newblock A survey of parallel and distributed algorithms for the steiner tree
  problem.
\newblock {\em International Journal of Parallel Programming'13}.

\bibitem{p:banks1}
G.~Bhalotia, A.~Hulgeri, C.~Nakhe, S.~Chakrabarti, and S.~Sudarshan.
\newblock Keyword searching and browsing in databases using banks.
\newblock In {\em Proceedings of International Conference on Data Engineering},
  ICDE '02.

\bibitem{p:lod}
C.~Bizer, T.~Heath, and T.~Berners-Lee.
\newblock Linked data-the story so far.
\newblock In {\em International Journal on Semantic Web and Information
  Systems}, IJSWIS'09.

\bibitem{p:JoelCoffmanValidation}
J.~Coffman and A.~C. Weaver.
\newblock A framework for evaluating database keyword search strategies.
\newblock In {\em Proceedings of the ACM international Conference on
  Information and Knowledge Management}, CIKM '10.

\bibitem{p:banks3}
B.~Dalvi, M.~Kshirsagar, and S.~Sudarshan.
\newblock Keyword search on external memory data graphs.
\newblock In {\em Proceedings of the international conference on Very Large
  Databases}, VLDB '08.

\bibitem{p:mr}
J.~Dean and S.~Ghemawat.
\newblock Mapreduce: Simplified data processing on large clusters.
\newblock In {\em Proceedings of symposium on Operating Systems Design and
  Implementation}, OSDI '04.

\bibitem{p:DPBF}
B.~Ding, J.~Xu~Yu, S.~Wang, L.~Qin, X.~Zhang, and X.~Lin.
\newblock Finding top-k min-cost connected trees in databases.
\newblock In {\em Proceedings of IEEE International Conference on Data
  Engineering}, ICDE'07.

\bibitem{p:PPI}
M.~T. Dittrich, G.~W. Klau, A.~Rosenwald, T.~Dandekar, and T.~M{\"u}ller.
\newblock Identifying functional modules in protein--protein interaction
  networks: an integrated exact approach.
\newblock {\em Bioinformatics}, 2008.

\bibitem{p:steiner}
S.~E. Dreyfus and R.~A. Wagner.
\newblock The steiner problem in graphs.
\newblock {\em Networks}, 1(3):195--207, 1971.

\bibitem{p:faTopK}
R.~Fagin.
\newblock Combining fuzzy information: an overview.
\newblock {\em SIGMOD Rec.}, 31(2):109--118, June 2002.

\bibitem{p:STPSmallK}
J.~Feldman and M.~Ruhl.
\newblock The directed steiner network problem is tractable for a constant
  number of terminals.
\newblock In {\em Proceedings of Annual Symposium on Foundations of Computer
  Science}, 1999.

\bibitem{p:xrank}
L.~Guo, F.~Shao, C.~Botev, and J.~Shanmugasundaram.
\newblock Xrank: Ranked keyword search over xml documents.
\newblock In {\em Proceedings of the ACM SIGMOD international conference on
  Management of Data}, SIGMOD '03.

\bibitem{p:kwdSrch:mr}
Y.~Hao, H.~Cao, Y.~Qi, C.~Hu, S.~Brahma, and J.~Han.
\newblock Efficient keyword search on graphs using mapreduce.
\newblock In {\em Proceedings of IEEE international conference on Big Data},
  BigData '15.

\bibitem{p:blinks}
H.~He, H.~Wang, J.~Yang, and P.~Yu.
\newblock Blinks: ranked keyword searches on graphs.
\newblock In {\em Proceedings of ACM SIGMOD international conference on
  Management of data}, SIGMOD '07.

\bibitem{p:Discover}
V.~Hristidis and Y.~Papakonstantinou.
\newblock Discover: keyword search in relational databases.
\newblock In {\em Proceedings of the international conference on Very Large
  Data Bases}, VLDB '02.

\bibitem{p:banks2}
V.~Kacholia, S.~Pandit, S.~Chakrabarti, S.~Sudarshan, R.~Desai, and
  H.~Karambelkar.
\newblock Bidirectional expansion for keyword search on graph databases.
\newblock In {\em Proceedings of the international conference on Very Large
  Databases}, VLDB '05.

\bibitem{p:rCliquesKS}
M.~Kargar and A.~An.
\newblock Keyword search in graphs: Finding r-cliques.
\newblock In {\em Proceedings of international conference on Very Large
  Databases}, VLDB '11.

\bibitem{p:sagiv1}
B.~Kimelfeld and Y.~Sagiv.
\newblock Finding and approximating top-k answers in keyword proximity search.
\newblock In {\em Proceedings of ACM Symposium on Principles of Database
  Systems}, PODS '06.

\bibitem{p:SocTeamSteiner}
T.~Lappas, K.~Liu, and E.~Terzi.
\newblock Finding a team of experts in social networks.
\newblock In {\em Proceedings of ACM SIGKDD international conference on
  Knowledge Discovery and Data mining}, KDD'09.

\bibitem{p:MannilaKeff}
T.~Lappas, E.~Terzi, D.~Gunopulos, and H.~Mannila.
\newblock Finding effectors in social networks.
\newblock In {\em Proceedings of ACM SIGKDD international conference on
  Knowledge Discovery and Data Mining}, KDD '10.

\bibitem{p:banks_ease}
G.~Li, B.~Ooi, J.~Feng, J.~Wang, and L.~Zhou.
\newblock Ease: an effective 3-in-1 keyword search method for unstructured,
  semi-structured and structured data.
\newblock In {\em Proceedings of ACM SIGMOD international conference on
  Management of Data}, SIGMOD '08.

\bibitem{p:pregel}
G.~Malewicz, M.~H. Austern, A.~J. Bik, J.~C. Dehnert, I.~Horn, N.~Leiser, and
  G.~Czajkowski.
\newblock Pregel: a system for large-scale graph processing.
\newblock In {\em Proceedings of ACM SIGMOD international conference on
  Management of Data}, SIGMOD'10.

\bibitem{p:parallelSTP}
G.~L. Presti, G.~L. Re, P.~Storniolo, and A.~Urso.
\newblock A grid enabled parallel hybrid genetic algorithm for spn.
\newblock In M.~Bubak, G.~Albada, P.~Sloot, and J.~Dongarra, editors, {\em
  Computational Science - ICCS '04}, Lecture Notes in Computer Science.

\bibitem{p:SteinerSoln2000}
G.~Robins and A.~Zelikovsky.
\newblock Improved steiner tree approximation in graphs.
\newblock In {\em Proceedings of annual ACM-SIAM Symposium On Discrete
  Algorithms}, SODA '00.

\bibitem{p:distSTP2}
G.~Singh and K.~Vellanki.
\newblock A distributed protocol for constructing multicast trees.
\newblock In {\em Proceedings of International Conference On Principles Of
  Distributed Systems}, OPODIS '98.

\bibitem{p:kwdSrchSurvey}
J.~X. Yu, L.~Qin, and L.~Chang.
\newblock Keyword search in relational databases: A survey.
\newblock {\em IEEE Data Eng. Bull}, 2010.

\bibitem{p:invertedIndexSurvey}
J.~Zobel and A.~Moffat.
\newblock Inverted files for text search engines.
\newblock {\em ACM Comput. Surv.}, 2006.

\end{thebibliography}
% vldb_sample.bib is the name of the Bibliography in this case
% You must have a proper ".bib" file
%  and remember to run:
% latex bibtex latex latex
% to resolve all references

%APPENDIX is optional.
% ****************** APPENDIX **************************************
% Example of an appendix; typically would start on a new page
%pagebreak
\begin{appendix}
%================================================
\section{Fagin's Algorithm}
\label{sec:fagin}
\textbf{\textit{Brief description of Fagin's algorithm}}: Fagin's algorithm \cite{p:faTopK} is about finding top-K objects under sorted access of the 
attributes of the objects. Let us assume that there is a set of many objects $R = \{O_1, O_2, \dots\}$, and every object has $m$ attributes, i.e., $O_i = \{x^i_1, x^i_2, ..., x^i_m\}$. These $m$ attributes of various objects can be accessed from individually sorted lists $L = \{L_1, \dots, L_m\}$. An aggregate function $f(x_1, \dots, x_m)$, is used for calculation of weight of the object $O_i$, using its attribute 
values. For example, students ($O_i$) in a course compete for the top-K positions by performing well in $m$ subjects. The teachers of all 
subjects prepare a list of students' marks in descending order, and send it to course coordinator. The course coordinator wants to 
identify the top-K best performing students, from these individually sorted list of marks in every subject. 
If the aggregate function is monotonic with respect to these $m$ attributes, and these lists are accessed in parallel then, 
according to Fagin, it is sufficient to access these lists sequentially until all attributes of at-least K objects are seen in these lists.
Further, in order to access these $K$ objects, $M$ objects would also have been seen but partially, i.e., $M$ objects
were seen in less than $m$ lists. The remaining attributes of these $M$ objects should then be accessed randomly. As a
result, $(K+M)$ Objects will be known. Fagin stated that the top-K objects according to their weights will be within these $\mathcal{C} = 
(K+M)$ objects, and therefore these are referred to as \textit{candidate top-K objects}. Here, a function is called monotonic if
following condition is satisfied: $f(x_1, \dots, x_m) \leq f(x'_1, \dots, x'_m)$, whenever $\forall i,~x_i \leq x'_i$. 
%================================================
\end{appendix}

\end{document}